# Reconfigurable radiofrequency electronic functions designed with 3D Smith Charts in Metal-Insulator-Transition Materials

Andrei A. Muller[1*], Alin Moldoveanu[2], Victor Asavei[2], Riyaz A. Khadar[1], Esther Sanabria-Codesal[3], Anna Krammer[1], Montserrat Fernandez-Bolaños[1], Matteo Cavalleri[1], Junrui Zhang[1], Emanuele Casu[1], Andreas Schuler[1], Adrian M. Ionescu[1]

Recently, the field of Metal-Insulator-Transition (MIT) materials has emerged as an unconventional solution for novel energy efficient electronic functions, such as steep slope subthermionic switches, neuromorphic hardware, reconfigurable radiofrequency functions, new types of sensors, teraherz and optoelectronic devices. Designing radiofrequency (RF) electronic circuits with a MIT material like vanadium dioxide, $VO_2$, requires the understanding of its physics and appropriate models and tools, with predictive capability over large range of frequency (1-100GHz).

In this work, we develop 3D Smith charts for devices and circuits having complex frequency dependences, like the ones resulting by the use of MIT materials. The novel foundation of a 3D Smith chart involves here the geometrical fundamental notions of *oriented curvature* and *variable homothety* in order to clarify first theoretical inconsistencies in Foster and Non Foster circuits, where the driving point impedances exhibit mixed clockwise and counter-clockwise frequency dependent paths on the Smith chart as frequency increases. We show here the unique visualization capability of a 3D Smith chart, which allows to quantify orientation over variable frequency. The new 3D Smith chart is applied as a 3D multi-parameter modelling and design environment for the complex case of Metal-Insulator-Transition (MIT) materials where their permittivity is dependent on the frequency.

In this work, we apply 3D Smith charts to on Vanadium Dioxide ($VO_2$) reconfigurable Peano inductors. We report fabricated inductors with record quality factors using $VO_2$ phase transition to program multiple tuning states, operating in the range 4 GHz to 10 GHz. Finally, the 3D Smith chart enables the design and fabrication of a new variety of Peano curves filters that are also used to experimentally extract the frequency-dependent dielectric constant of $VO_2$ within 1 GHz- 50 GHz, a quasi-unexplored field of major importance for the accurate predictive design of radio frequency electronic applications with phase change materials.

The Smith chart, invented in 1939[1], is a graphical tool widely used in various fields of electrical engineering and applied physics when dealing with frequency dependent reflection coefficients or impedances. The Smith chart is widely employed in the design/measurement stage of a large variety of circuits, from metasurfaces[2] to coils[3] (Supplementary Fig. 3 in[3]) or scanning microwave microscopy[4], while being mostly present in microwave-terahertz frequency region in the design and characterization of antennas[5], transmission lines[6-7], power amplifiers[8], filters[9] or acoustic resonators[10]. The 3D Smith chart proposed in[11] generalizes the Smith chart (which is limited within the unit circle to circuits with reflection coefficients ($\Gamma = \Gamma_r + j\Gamma_i$) magnitudes smaller than unity Fig. 1a) onto the Riemann sphere in order to make it usable for all circuits (Fig. 1b and Fig. 1c).

An essential drawback of the Smith chart and previous 3D Smith chart representations[1-13] is the lack of visualization of the variable parameter (frequency), thus the orientation changes and dynamics of the scattering parameters frequency dependency being impossible to be pictured. Although many circuits exhibit a clockwise orientation of their driving point impedances and reflection parameters curves as frequency increases[2-5,9],([6,10] – unspecified), the absence of a clockwise motion (i.e. discontinuity points or counter-clockwise motion) was often reported leading to diverse interpretations. In active devices (transistors), as for example in[14,15] it is referred as "kink-phenomenon", in lossless (purely reactive) non-Foster circuits (such as negative capacitors and inductors)[16-19], as an intrinsic phenomenon. Unfortunately, this counter-clockwise dynamics phenomenon recurrent existence in lossy circuits with non-Foster elements[18,20-23], led to some misleading conclusions: in[20-23] the authors assume

1 | Ecole Polytechnique Federale de Lausanne, 1015 Lausanne, Switzerland
2 | University POLITEHNICA of Bucharest, 060042 Bucharest, Romania
3 | Universitat Politècnica de València, 46022 Valencia, Spain
* | Author to whom correspondence should be addressed: andrei.muller@epfl.ch

that the existence of this phenomenon proves the presence of a non-Foster element, while in[24] it is stated that passive linear devices cannot exhibit driving point immittances with counter-clockwise frequency dependency on the Smith chart.

2D Smith chart representations lacking in zooming may oversee the changes of orientation occurring in passive networks (for input impedances and reflection coefficients) with Foster elements (observed too in[25-26]). This orientation reversal phenomenon in lossy networks with Foster elements is often overlooked or seen an interpolation error and its presence is often ignored. Since the paths of the reflection coefficients of these circuits (excepting orientation) may be identical the Smith chart on a specific frequency range (as for negative inductors and positive capacitors and viceversa[16,18,19,25]) a frequency dependency visualization is missing in order to get an insight to the intrinsic phenomenon.

For this purpose, we first introduce the notion of oriented curvature[27], prove the mixed clockwise-counter-clockwise orientation phenomenon in lossy circuits with Foster elements and propose a frequency orientation quantification (while increasing the sweeping range) with a new implementation within the 3D Smith chart tool by using the topology of the Riemann sphere[28-29]. Thus, a new vision to detect this orientation reversal for both Foster and Non-Foster[16-23,30-32] elements based networks is first presented (while increasing the frequency of analysis), impossible to quantify for 2D parametrical plot such as the Smith chart or basic 3D Smith chart (where only the Riemann sphere surface is used).

Further the 3D Smith chart is additionally exploited and further developed in order to display simultaneously parameters needed in reconfigurable frequency designs while dealing with inductors. Frequency dependent inductances based on the $Y_{11}$[33-35] admittance parameter (shunt models), series models of inductances based on the $Y_{21}$[36-37] parameter, quality factors[33-38], self-resonant frequencies, Smith chart information[38] need all to be simultaneously optimized and analyzed over a wide frequency range during the designs and characterizing stages of inductors. This leads[33-38] to a larger number of graphics or to different scaling in order to grasp all parameters of interest. Here all these factors are implemented and analyzed on a display using the 3D Smith chart topology and different perspectives.

The newly developed capabilities are particularly suited to explore and design reconfigurable CMOS-compatible inductors design for microwaves frequencies using a phase change material like Vanadium Dioxide ($VO_2$)[36-37, 39-46] for tuning the values of inductance. Indeed, it is known that $VO_2$ behaves like an insulator under its phase transition temperature Tc = 68 °C (or higher when dopped[41]) with monoclinic crystal structure[39] while deposited on $SiO_2$/Si substrates. Because of its ease of integration, reversible insulator to metal transition (IMT), low transition temperature and fast switching time, the employment of Vanadium Dioxide ($VO_2$) as a reconfigurable radio frequency (RF) material has been just recently investigated for a variety of RF-reconfigurable devices[36-37,39,43,44]. Still, much of existing studies are carried out in the frequency range of terahertz or far-infrared[40,42-43], leaving (RF) $VO_2$ a largely uncharted area for exploration in development. The conductivity levels of $VO_2$ in its off (isolating state) and on state (conductive state) vary over a wide range depending on the substrate[36-37,39-46] causing limitations in the RF devices performances (being below 50,000S/m for $SiO_2$/Si depositions in the on state).

The on state limited conductivity levels restricted the quality factors of the reconfigurable inductors fabricated with $VO_2$ in its conductive state[37] to sub-unitary values, while to less than three in[36] for CMOS compatible Si $VO_2$ reconfigurable inductors.

Here we further introduce a new type of reconfigurable inductors based on the Peano curves[47] with $VO_2$ switches and using the freshly implemented multi parameter displays we manage to increase with a factor of 2.33 the quality factors previously reported in[36] at reconfigurable inductors with $VO_2$ in on state (and with orders of magnitude in respect to[37]) while increasing too the number of tuning states. Further due to its original geometry the inductor exhibits a tuning range of 77% (better than 55% in[36] or 32% in[37]) and a ratio of $Q_{max\_on}$/ $Q_{max\_off}$ of 0.87 unlike 0.27 in our previous work[36] on $SiO_2$/Si substrates. It is worth pointing out that the inductor while facing the limited conductivity levels of $VO_2$ on $SiO_2$/Si substrates exhibits a 2.33 better quality factor in the on state also in respect to other $VO_2$ based reconfigurable inductors such as the $SiO_2$/Sapphire inductor reported in[48] where the conductivity levels exceed 300.000S/m.

Motivated by the large variety of the values reported at room temperatures for the dielectric constant[40] (the real part of the relative effective permittivity) of the $VO_2$ in its semiconductor state (30 in[36,44](stated) or 90-700 in the 0.1 GHz-30 GHz range (frequency dependent[43]), 3-10 at infrared frequencies[40,45]) we elaborate and validate a new model of the effective permittivity of $VO_2$. The model in[43] is facing the limitations of the conformal mapping extraction techniques[46] while dealing with finite ground planes (assumed infinitely in[43]), skin effect losses[49] while neglecting damping losses[50] in the final extraction model of the relative effective permittivity.

By producing a series of Peano inspired filters (using the space filling property of the curves to shift resonance frequencies down while using limited areas) with a series of different resonance frequencies with or without a thin $VO_2$ layer below them we propose based on a mixed frequency shift model and avoiding the limitations of conformal mapping[43] a more complete model of the effective permittivity including both intrinsic dielectric losses (damping) and conductive losses[50] and we extract the frequency dependency of the dielectric constant. The extraction is performed within the commercial, satellites communications frequency bands, from 1 GHz-50 GHz and thus provide a more complete model for this emerging material whose applications in reconfigurable wireless communications are in its birth moments.

**Oriented curvature of input impedances, reflection coefficients, slope of reactance and 3D Smith chart implementation of frequency dependency orientation**

Based on the notion of oriented curvature introduced and explained in the Supplementary Material, we show that the changes in sign of the reactance frequency derivative do not imply changes in orientation neither for the input impedance nor for the reflection coefficient of 1- port networks (when losses occur). We provide the conditions and equations under which one may have the same orientation (more details in Supplementary Section 1) for both reflection coefficient and input impedances. The lossless (reactive) cases (purely Foster[16-17,26] and non-Foster[16-19]) become particular cases where the reflection coefficients are direct inversive (Mobius) and indirect inversive transformations of the oriented imaginary axes of the impedance plane and the clockwise and counter-clockwise motion on circles is a consequence of the reactance slope and sign.

By introducing the geometrical notion of oriented curvature in this field[27] we prove that the assumptions made by other authors[20-24] may not apply (see Supplementary Section 1). Further, as seen also for the input impedance of an antenna in[26], the negative frequency derivative of the reactance of a lossy 1-port network does not imply counter-clockwise motions in the case of lossy 1-port networks.

Consider first a 1-port network terminated on a resistive load $r$. The input impedance is given by (1), where $r_m(\omega)$ denotes its resistive part and $x_m(\omega)$ its reactive part, while its reflection coefficient is given by (2). Computing the oriented curvature values for both of them (Supplementary Section 1) we get $k_{zm}(\omega)$ (the oriented curvature of the input impedance) and $k_{\Gamma_{1zm}}(\omega)$ (the oriented curvature of the 1-port reflection coefficient) as (3) and (4).

$$z_m(j\omega) = r_m(\omega) + jx_m(\omega) \tag{1}$$

$$\Gamma_{1zm}(j\omega) = \frac{z_m(j\omega)/r - 1}{z_m(j\omega)/r + 1} \tag{2}$$

$$k_{zm}(\omega) = \frac{r_m'(\omega)^2 \left(\frac{x_m'(\omega)}{r_m'(\omega)}\right)'}{(r_m'(\omega)^2 + x_m'(\omega)^2)^{3/2}} \tag{3}$$

$$k_{\Gamma_{1zm}}(\omega) = \frac{r_m'(\omega)^2((r+r_m(\omega))^2 + x_m(\omega)^2)\left(\frac{x_m'(\omega)}{r_m'(\omega)}\right)' + 2x_m(\omega)^2(r_m'(\omega)^2 + x_m'(\omega)^2)\left(\frac{r+r_m(\omega)}{x_m(\omega)}\right)'}{2r(r_m'(\omega)^2 + x_m'(\omega)^2)^{3/2}} \tag{4}$$

(i) If $k_{zm}(\omega) > 0$ then $z_m(j\omega)$ is counter-clockwise oriented; (ii) if $k_{zm}(\omega) < 0$ then $z_m(j\omega)$ is clockwise oriented; (iii) if $k_{\Gamma_{1zm}}(\omega) > 0$ then $\Gamma_{1zm}(j\omega)$ is counter-clockwise oriented; (iv) $k_{\Gamma_{1zm}}(\omega) < 0$ then $\Gamma_{1zm}(j\omega)$ is clockwise oriented; (v) If $k_{zm}(\omega) = 0$ then $z_m(j\omega)$ is a line; (vi) If $k_{\Gamma_{1zm}}(\omega) = 0$ then $\Gamma_{1zm}(j\omega)$ is a line. The complex link between the change of sign of (3) and (4) is described in Supplementary Section 1.

Denoting with $x_{mF}(\omega)$ and $B_{mF}(\omega)$ the reactance and susceptance of purely Foster elements and with $x_{mNF}(\omega)$ and $B_{mNF}(\omega)$ the ones for Non Foster elements we obtain: for Foster networks $r_m(\omega) = 0$ while $\frac{dx_{mF}(\omega)}{d\omega} > 0$ and $\frac{dB_{mF}(\omega)}{d\omega} > 0$ and using (3)-(4) we get the input impedance and 1–port reflection coefficient curvatures for them as: $k_{Fzm}(\omega) = 0$ and $k_{\Gamma_{1zmF}}(\omega) = -1$ (Supplementary Section 1). For Non-Foster networks $r_m(\omega) = 0$ too, while $\frac{dx_{mNF}(\omega)}{d\omega} < 0$ and $\frac{dB_{mNF}(\omega)}{d\omega} < 0$ hold thus via (3) and (4) we get the input impedance and 1–port reflection coefficient curvatures for them as $k_{NFzm}(\omega) = 0$ and $k_{\Gamma_{1zmNF}}(\omega) = 1$.

In the case of two port networks with equal port impedances similar computations can be done for purely Foster and non-Foster elements resulting in the corresponding reflection coefficients $\Gamma_{2zmF}$ and $\Gamma_{2zmNF}$ with their corresponding oriented curvatures $k_{\Gamma_{2zmF}}(\omega) = -2$ and $k_{\Gamma_{2zmNF}}(\omega) = 2$. Their oriented curvature magnitudes explain their paths on 0.5 radius circles in Fig. 1d and Fig. 1e on the Smith chart and 3D Smith chart) (see Supplementary

Section 1). The reflection coefficients on the other hand of purely reactive 1-port elements are given in Fig. 2a-Fig. 2d (purely reactive Foster and non-Foster circuits in Fig. 2a, lossy circuits with non-Foster elements in c and lossy circuits with Foster elements in Fig. 2d. The results in Fig. 2d show that clockwise and counter clockwise reflection coefficients rotations (and the same for input impedance) can occur at lossy 1-port networks containing only Foster elements too.

The results plotted in Fig. 1d and Fig. 1e, Fig. 2b - Fig. 2c show the new 3D Smith chart implementations capable of detecting orientation changes and intrinsic behavior where the Smith chart shows a limited insight capability to the phenomenon. Supplementary Section 2 describes fully the employment details.

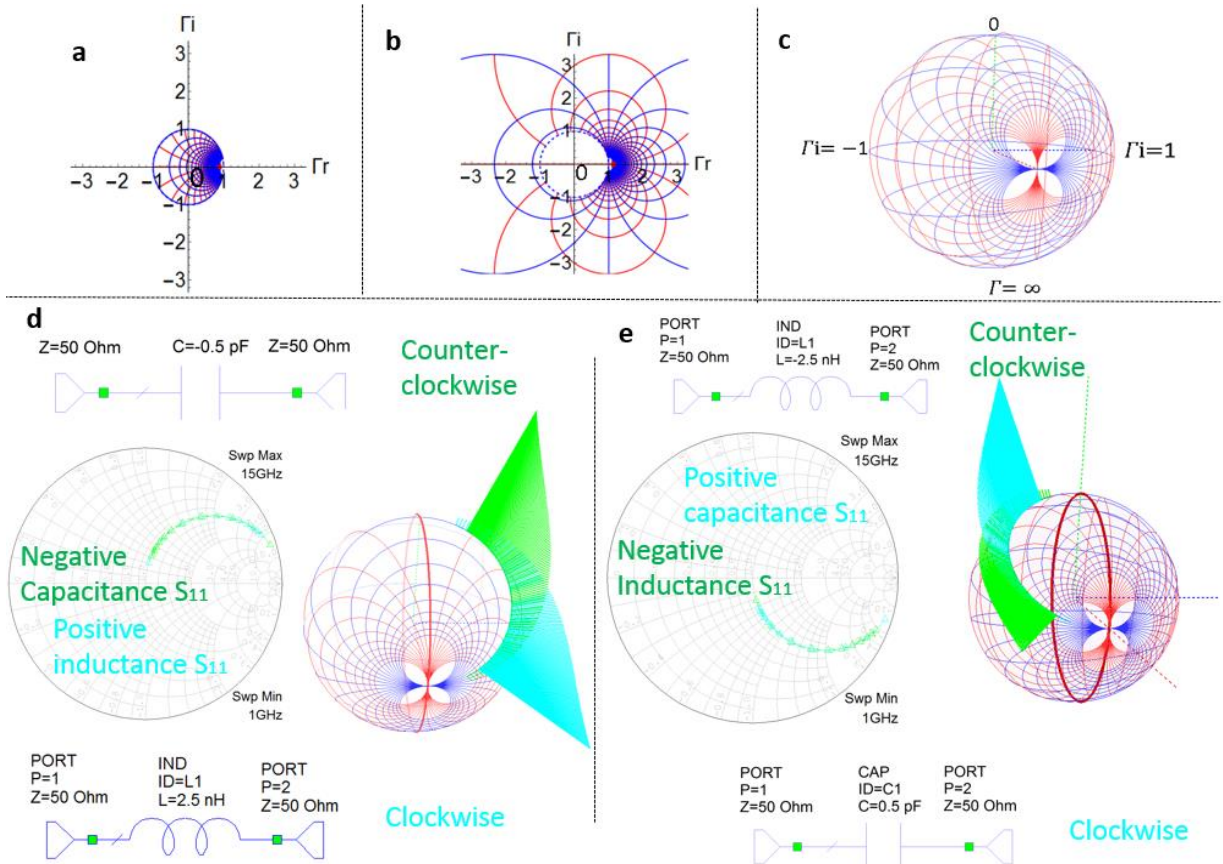

**Fig. 1| Smith chart limitations and clockwise and counter-clockwise frequency dependency of Foster and non-Foster elements on a newly introduced frequency dependent 3D Smith chart. a,** Smith chart. **b,** Extended Smith chart **c,** 3D Smith chart as in 2018 (without frequency dependency). **d, 3D** Smith chart representation of the two port reflection coefficient and newly implemented frequency dependent 3D Smith chart representation for it. For a capacitor with purely negative capacitance and an inductor with positive inductance their reflection coefficient frequency representation overlaps on the Smith chart for a wide frequency range, their orientation changes cannot be distinguished. On the newly implemented frequency dependent 3D Smith chart one can clearly see their clockwise motion with increasing frequency for the inductor while the counter-clockwise motion for the capacitor with negative capacitance. **e,** 3D Smith chart representation of the reflection coefficient and newly implemented frequency dependent 3D Smith chart for a negative valued inductor and capacitor with positive capacitance. Again, their trace is identical on a Smith chart, their intrinsic opposed frequency dependency cannot be seen. On the 3D Smith chart one can directly see the clockwise motion of the reflection coefficient of the capacitor and the counter-clockwise motion of the reflection coefficient of the negative valued inductor.

The main new insight is given by the representation of the frequency parameter over the 3D Smith chart representation of the reflection coefficient $S_{11_{3d}}(j\omega)$ via a variable homothety with its center in the center of the 3D Smith chart: Each frequency that corresponds to a point of the 3D Smith chart reflection point of the $S_{11_{3d}}(j\omega)$ curve will be displayed as a segment on the line that passes from the center of the 3D sphere and the point of the 3D Smith chart surface curve of $S_{11_{3d}}(j\omega)$. The length of the segment will be given by the normalized frequency and the direction will be outwards of the surface of the 3D sphere. Fig 1d displays in the 3D Smith chart surrounding space the counter-clockwise dynamics of the two port negative capacitor reflection coefficient while the clockwise dynamics of the two port reflection coefficient of the positive inductor. In Fig. 1e one may see the clockwise frequency increasing orientation of the reflection coefficient of the positive capacitor and the counter-clockwise orientation of the negative inductor.

The Smith chart plot can detect the magnitude of the curvature $|k_{\Gamma 2F}(\omega)|$ (which gives the path of the reflection coefficient) but cannot see its sign which determines its direction, the new 3D Smith chart orientation quantification implementation (the frequency sweeping is always increasing in our modelling) detects its sign (see additional video) and thus its orientation. In the cases presented in Fig. 1d and Fig. 1e $|k_{\Gamma 2F}(\omega)|$ is constant and not zero thus a circle. The same happens in Fig. 2a and Fig. 2b. However in more complex circuits one does not deal with constant curvatures, these can alternate in sign and frequency dependency exhibiting orientation changes for both Foster and Non Foster circuits as seen in Fig. 2c and Fig. 2d. In Fig. 2d it can be seen that even a network characterized by a positive real function can generate mixed curvature in its input impedance and reflection coefficient. These reversals of orientation may be easy overlooked on the Smith chart if the zooming scales are not properly chosen but using (3) and (4) this is clearly discovered in Fig. 2d.

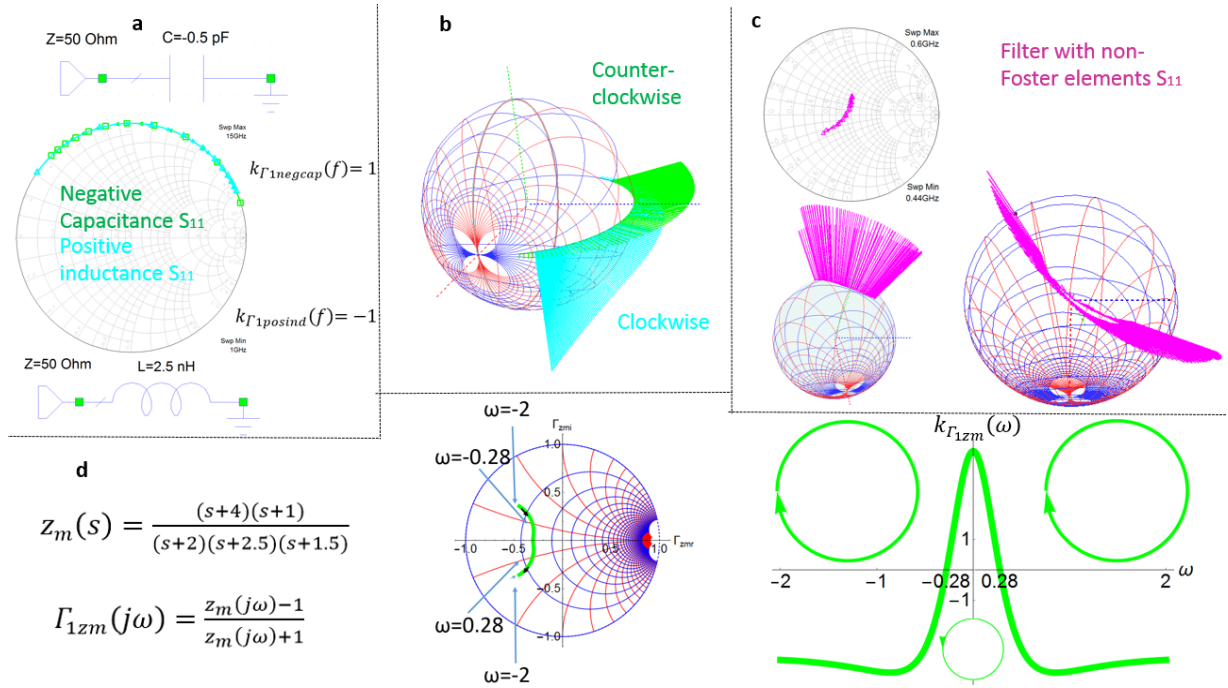

**Fig. 2| Reflection coefficient orientation changes and the sign of the oriented curvature for different circuits. a,** Smith chart representation of the reflection coefficient for a 1-port negative capacitance (purely non-Foster circuit) and a positive inductance (purely Foster). For a capacitor with purely negative capacitance and an inductor with positive inductance their reflection coefficients $\Gamma_{1zm}(j\omega)$ overlap on the Smith chart on a wide frequency range. Their opposite orientation is given by the different sign of their oriented curvature $k_{\Gamma_{1zm}}(\omega)$. Their same path is given by the same absolute value of their oriented curvature. **b,** On the newly implemented frequency dependent 3D Smith chart one can see the clockwise motion with increasing frequency for the inductor with positive inductance and the counter-clockwise motion for the negative valued capacitor, the motion is on the contour of the equatorial plane (lossless circuits). **c,** Mixed motion for a fabricated circuit containing non-Foster (lossy elements). **d,** Mixed clockwise and counter-clockwise motion of the reflection coefficient of a passive lossy network described by the positive real function $z_m(s)$ with the 1 port reflection coefficient (for s=j$\omega$) $\Gamma_{1zm}(j\omega)$. The reflection coefficient has a clockwise orientation from -2< $\omega$<-0.28 and for 0.28< $\omega$<2, while counter-clockwise orientation for -0.28< $\omega$<0.28. The sign changes of its 1-port reflection coefficient curvature $k_{\Gamma_{1zm}}(\omega)$(4) generates the changes of orientation of its path on the Smith chart. It is interesting to notice that mixed motion can exist on limited bandwidth also for lossy circuits with only Foster elements and thus that the counter-clockwise motion is by no means a prove of an existence of a non-Foster element in the network. A more detailed description on oriented curvature and 1-port and two port networks is given in Supplementary Section 1.

**3D visualization insights of frequency dependent series and shunt inductances and quality factors**

The S parameters of the inductors are directly converted by the new implementations in the software tool into the series inductance model $LSeries(\omega)$[36-37] and shunt inductance model $LShunt(\omega)$[33-35] using classical conversion techniques of two port parameters (see Supplementary Section 2). The series and shunt inductances values are then normalized to their maximum value over the frequency range of interest and we get the corresponding normalized values $LSeries_N(\omega)$ and $LShunt_N(\omega)$. The reflection parameter $S_{11}(j\omega)$ of the inductor is then plotted first on the surface of the 3D Smith chart as $S_{11_{3d}}(j\omega)$ using the previously implemented features[11-12]. Then the 3D space surrounding the 3D Smith chart is used by means of a variable homothethy with the homothetic center in the center of the sphere through the $S_{11_{3d}}(j\omega)$ parameter of the inductors. The $S_{11_{3d}}(j\omega)$ parameter is sent to another point in 3D at a distance corresponding to $LSeriesN$ or $Lshunt_N$:

$$LSeries, shunt_{3d}(\omega) = (LSeriesN, shunt_N(\omega) + 1) * S_{11_{3d}}(j\omega) \qquad (5)$$

The quality factor $Q(\omega)$ of the inductors is also computed by the new 3D Smith chart tool implementation using classical conversion formulas (from the S parameters) (see Supplementary Section 2) and normalized to its maximum over the frequency range of interest obtaining $Q_N(\omega)$. Using the 3D representation of the $LSeries, shunt_{3d}(\omega)$ curves we then use the normal plane of the curves to associate to each (frequency) point of them the quality factor as a cylinder of variable radius associated to its normalized value $Q_{3D}(\omega)$. In Fig. 3 one may see the implementations done for the newly fabricated Peano inductor compared whose performances are compared with the spiral inductor in[36].

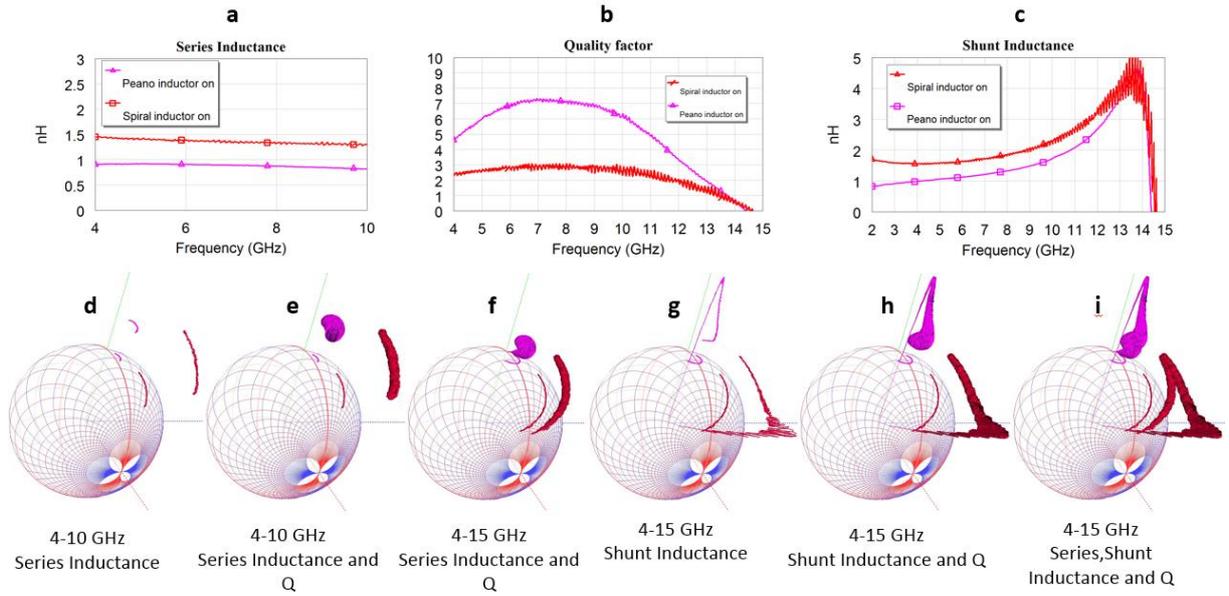

**Fig. 3| Series, shunt and quality factor frequency variations representations on a 2D plot and simultaneous 3D Smith chart visualization for the new fabricated Peano (pink) inductor in the conductive state of VO₂ versus previously reported best performing inductor (red) using VO₂ in same technology[36] a,** 2D series inductance representation versus frequency: both Peano inductor and previously reported inductor[36] show stable series inductance in within the 4GHz-10GHz frequency range. **b,** 2D frequency dependency of quality factors in the 4 GHz-10 GHz frequency range. **c,** Shunt inductance frequency dependency shows linearity on a sharper frequency range (see Supplementary Section 3) for both inductors. **d,** 3D Smith chart representation of series inductance over the $S_{11}$ parameters in the 4 GHz-10 GHz frequency range. **e,** 3D Smith chart representation of series inductance and quality factors 4GHz-10GHz over the series inductances 3D curves. **f,** 3D Smith chart representation of series inductance and quality factors 4GHz-15GHz. The Qs start descending to 0 close to 15 GHz (the cylinders radius becomes 0 when Q becomes negative) for both models while the $S_{11}$ parameters crossed into the West hemisphere of the Smith chart (capacitive region). **g,** 3D Smith chart representation of shunt inductance 4GHz-15GHz. One may see that the shunt inductance becomes negative below 15GHz (entering the 3D Smith chart) .**h**, 3D Smith chart representation of shunt inductance and quality factors 4GHz-15GHz.**i**, Simultaneously 3D Smith chart representation of series, shunt inductance and quality factors (along both series and shunt inductance)in the 4GHz-15GHz range.

The new implementation allows the concurrent view of five different parameters: $S_{11_{3d}}(j\omega)$, $LSeries_{3d}(\omega)$, $LShunt_{3d}(\omega)$, $Q_{3D}(\omega)$ and frequency (not plotted in Fig. 3 since the dynamics of $S_{11}(j\omega)$ is clockwise anyway in this case, unlike the cases presented in Fig. 2). The use of different perspectives and the topology of the 3D Smith chart permits one thus to simultaneously analyze matching problems (Smith chart) and visualize series and shunt inductances and quality factors all on a same interactive display. This plays an important role in investigation for directly understanding multiple phenomenon on a single display. The information contained in Fig. 3a-Fig.3c can be visualized together using three scaling on a common 2D plot, however still without having any information on the complex reflection coefficient $S_{11}(j\omega)$ of the inductor. In Fig. 3d we may see just $S_{11_{3d}}(j\omega)$ and $LSeries_{3d}(\omega)$, the display contains already more information than in the 2D Fig. 3a, allowing us to understand that the series inductance model is linear for both analyzed inductors and that $S_{11_{3d}}(j\omega)$ is still in the East hemisphere (in the 4GHz-10 GHz frequency range) (inductive). Additionally, the zeros of the $S_{11}(j\omega)$ are strongly related to the zeros of the $Y_{11}(j\omega)$ of an inductor and under certain circumstances (in some) identical (no resistive losses in their equivalent Pi model), thus a change of hemisphere of the $S_{11}(j\omega)$ is strongly related to the self-resonances of the inductors model implying in most cases that the quality factor fails to be positive anymore (see Supplementary Section 3). In Fig. 3e we can see $S_{11_{3d}}(j\omega)$, $LSeries_{3d}(\omega)$ and $Q_{3D}(\omega)$ for the 4 GHz-10 GHz frequency range, again the information contained offers an insight on three parameters impossible to visualize together in 2D. In Fig. 3f we may see how by increasing the analysis range up to 15 GHz the quality factors for both inductors become zero, while the $S_{11_{3d}}(j\omega)$ enters the capacitive hemisphere too. It is interesting to notice that the series inductance model stays linear. Using the 3D Smith chart tool (see supplementary video) one may see the exact frequencies at which $S_{11_{3d}}(j\omega)$ changes hemisphere and the frequency for which Q becomes zero (the 3D generalized cylinders become curves-their radius becomes zero (at 14.44 GHz for both inductors). In Fig. 3g $S_{11_{3d}}(j\omega)$, $LShunt_{3d}(\omega)$ models become negative starting 14.44 GHz for both inductors and thus they enter the interior of the 3D Smith chart from that frequency point. In Fig. 3h we may see $S_{11_{3d}}(j\omega)$, $LShunt_{3d}(\omega)$ and $Q_{3D}(\omega)$, clearly $Q_{3D}(\omega)$ becomes zero once $LShunt_{3d}(\omega)$ enters the 3D Smith chart. In Fig. 3i all these parameters are shown together.

The new implementation helps one to visualize key parameters of interest in the same time and thus supports and gives new insights, unreachable by any other means to our best knowledge, in the design stages of an inductor where one needs Q↑, stable series and shunt inductance (or both) over a large frequency range while maintaining the matching information $S_{11_{3d}}(j\omega)$.

**Peano reconfigurable inductors using VO$_2$ switches**

As a case study of the 3D Smith chart and its predictive capabilities for radiofrequency design with VO$_2$, we have designed and fabricated a reconfigurable inductor based on the Peano curve of order 2 by means of VO$_2$ switches. Its geometry is presented in Fig. 4. In Fig. 4a one may see the cross sectional view of the technology used. The inductors were fabricated using standard microelectronic processes starting with a high-resistivity (10000 Ω·cm) 525 μm thick silicon substrate. A 300 nm thick amorphous silicon layer was first deposited to improve radiofrequency performances. The substrate was then passivated with 500 nm SiO$_2$ deposited by sputtering. 140 nm-thick VO$_2$ and films a Pulsed Laser Deposition (PLD). The film was then patterned using photolithography followed by dry etching and Cr (20 nm)/Al (400 nm) bi-layer was deposited to contact the patterned VO$_2$ film. This thin contact layer allowed for the realization of smaller than 0.6 μm gaps between the contact pads (unlike 2μm in our previous work[36]). Additionally, a 2.4μm-thick Al layer was deposited on top of these contact pads by conventional lift-off methods to provide low RF losses (to create the final CPW elements), the photo of the fabricated inductor being shown in Fig. 4b.

The small width of the VO$_2$ switches (0.6 μm at the limit of photolithography) minimizes the losses while in the on state, while their increased length (120 μm) contributes too to this effort (a tilted photo of the switch in Fig. 4b, is in detail in (Supplementary Section 3). Switch photos and current distributions are further shown in Fig. 4c- Fig. 4f.

The inductor has been designed and simulated in the Ansys HFSS commercial software tool with the aim: to obtain a higher Q than in[36] while facing the limited VO$_2$ conductivity in the on state, measured as 40,000S/m), $Q_{max\_on}$/$Q_{max\_off}$ ↑ and more tuning range than 55%. The position of the VO$_2$ switches was optimized in order to maximize these values: the final current distribution at 5 GHz being shown as simulated in Fig. 4c (when both switches are off), Fig. 4e when one switch is off and one on and Fig. 4f with both switches on.

The measured inductances and quality factors plotted in Fig. 3a- Fig. 3b are compared to our previous work[36] for the on state of the inductor with two switches show ("on" when measured at 100°C): Peano inductors more than double the quality factor while also dealing with a smaller inductance (0.9 nH unlike 1.35 nH in[36], while usually the $Q_{max}$ decrease a lot while using lower inductances[34]) for the 5GHz-10 GHz frequency range. Further the series inductance is stable with an average value of 0.95 nH within the 4GHz-10 GHz frequency range and thus over-performing our previous reported results[36]. In terms of shunt inductance (untreated in[36,37]) the inductor is stable within the 3 GHz-6 GHz frequency range with an average of 1nH as seen in Fig. 3c.

The overall performances of the inductor with two switches on, two switches off and of a fabricated inductor with one switch on are all plotted together in Fig. 4g (series inductances), Fig. 4h quality factors and Fig. 4i both quality factors and normalized 3D series inductances and quality factors. The results show 77% tuning range and $Q_{max\_on}$/$Q_{max\_off}$ 0.88 and thus 3.26 times higher than $Q_{max\_on}$/ $Q_{max\_off}$ reported in[36]. The maximum quality factor in the on state exceeds 7 being comparable with the off state $Q_{max}$ even though the inductance is tuned with 77% down to 0.95 nH. This result in the on state is particularly interesting since it overcomes the limited performances of the previously reported inductor on $VO_2$[36] while using a smaller inductance (0.95nH vs 1.35nH).

On the other hand, the limited off state (at room temperature) quality factor (Q) is comparable with the one reported in[36] (although dealing with a smaller inductance than in[36]). While exhibiting a better performance in the low GHz frequency range and a more stable frequency dependency linearity the values are facing the same trend as in[36]. The maximum value is limited as in[36] by the CMOS compatible CPW $SiO_2$/Si technology used, (Supplementary Section 3). Additionally, the $VO_2$ lossy dielectric behavior in the off state, DC conductivity (20S/m) and intrinsic losses contribute to this. It is worth mentioning that comparable values of the Q were obtained here for smaller inductances

The design of radiofrequency functions in wideband applications, with accurate predictive capability, pointed out the need of an accurate modeling and extraction of main parameters of $VO_2$ such as the relative effective dielectric permittivity as a function of frequency, for which the reports in the literature are limited or inconsistent. A new extraction approach and new data are presented in the following section.

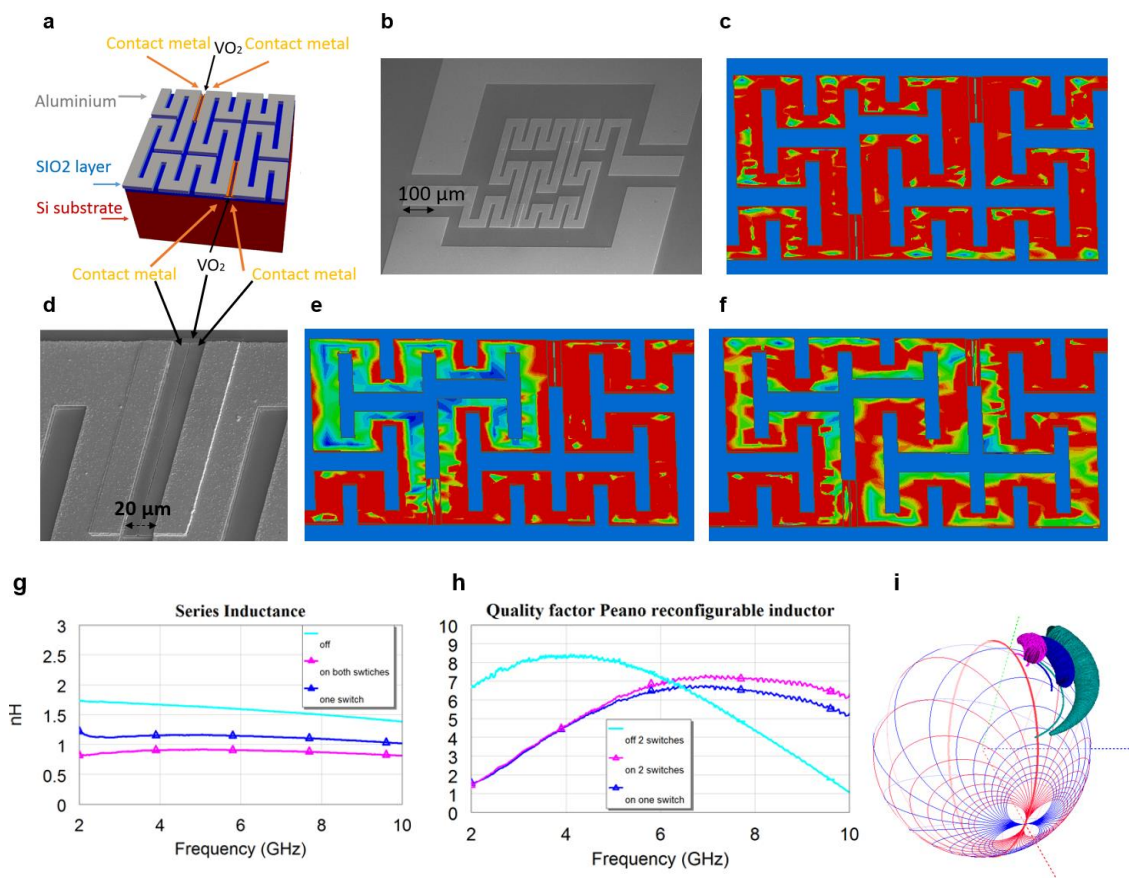

**Fig. 4| Fabricated VO$_2$ reconfigurable Peano inductor geometry and performances a,** Cross sectional view **b,** SEM photo of the inductors. **c,** Current distribution ( 5GHz) in the off state for an inductor with 2 switches. The current ignores the VO$_2$ switches and goes through all the windings in its path from port 1 (input) to port 2 (output). **d,** VO$_2$ switch fabricated photo: a gap in the contact metal of 0.6µm is left in order to contact the VO$_2$ layer. In the insulating phase of VO$_2$ the switch is in off state (acting like a lossy dielectric), in the conductive state of the VO$_2$ the 600nm gap plays an important role to minimize the conductive losses since the VO$_2$ has a limited 40,000S/m conductivity on the SiO$_2$/Si substrate. **e,** Current distribution (5GHz) in the hypothetical on state of one switch while the other is in the off state. The current ignores the off state VO$_2$ switch and follows a shorter path from input to output. **f,** Current distribution (5 GHz) in the on state for an inductor with 2 switches. All current distributions are within a range from 0.5 A/m (dark blue) to 5.2 A/m (intense red) using the same scale. **g,** Series inductance of the fabricated inductors with 2 switches, in on state and off state, series inductance of an inductor with one switch in on state. **h,** Quality factor of the inductor with 2 switches in off state and on state, and of an inductor with one switch (on state) **i,** Simultaneous 3D Smith chart representation for the 2 GHz-10 GHz frequency of the series inductance and quality factors for all 3 situations.

**Dielectric constant of VO₂ in the microwave frequency range**

Space filling curves can be also used to generate resonating structures with a variety of central frequencies using very limited space. To generate resonating frequencies within the 1 GHz- 50 GHz microwave spectrum we use tenths of Peano inspired defected ground plane structures (DGS) in the coplanar waveguides (CPW). Peano curves of order 3 kind of DGS (which have 9 times the complexity of the Peano curve of order 2) are used to generate resonance frequencies below 10 GHz (Fig. 5a), while modified repetitions of DGS Peano curves inspired shapes of order 2 are used to generate structures resonating within the 10 GHz-50 GHz frequency range (Fig. 5a). The structures have different widths of the tubes along the generating Peano types of curves and different aspect ratios. The structures are fabricated on bilayer SiO₂/Si (Fig. 5b) substrates and on multilayer VO₂/ SiO₂/Si substrates (Fig. 5c). Each of the filters performances is measured in both cases and a frequency shift and losses difference is noticed. The exact same DGS structure on the triple-layer substrate in Fig. 5c would always resonate at a lower frequency than on the bi-layer structure if the dielectric constant $\varepsilon'_{reff}(f)$ of VO₂ is higher. ($\varepsilon_{reff}(f)$ is the relative effective dielectric permittivity, $\varepsilon'_{reff}(f)$ its real part and $\varepsilon''_{reff}(f)$ its imaginary part while $\varepsilon_0$ the permittivity of the free space)-(6).However for each DGS type of structure this relative down shift level is lower or higher for a fixed dielectric constant of the VO₂, the relative shift itself being dependent also on each DGS type. The bandstop performances in terms of Q and maximum attenuation are on the other hand very sensitive to the losses of the VO₂ thin film. (see Supplementary Section 4)

To overcome limitations of previously reported models of the VO₂ of the $\varepsilon_{reff}(f)$ for the 1 GHz-50 GHz frequency ranges[43] we include the damping losses (within the first imaginary term in (6) using the more general model[50]) - the intrinsic tangent loss $tan\delta$ and separate this term from the losses given by the conductivity $\sigma'(f)$. The presence of this term enables us to perform the de-embedding with HFSS simulator[50] which allows the user to make a distinction between conductive losses determined by the non-zero conductivity $\sigma'(f)$ of the VO₂ and other intrinsic damping losses of the dielectric included in $\delta$.

$$\varepsilon_{eff}(f) = \varepsilon_0 * \varepsilon'_{reff}(f) * (1 - jtan\delta - j\sigma'(f)/(\omega\varepsilon_0\varepsilon'_{reff}(f))) \qquad (6)$$

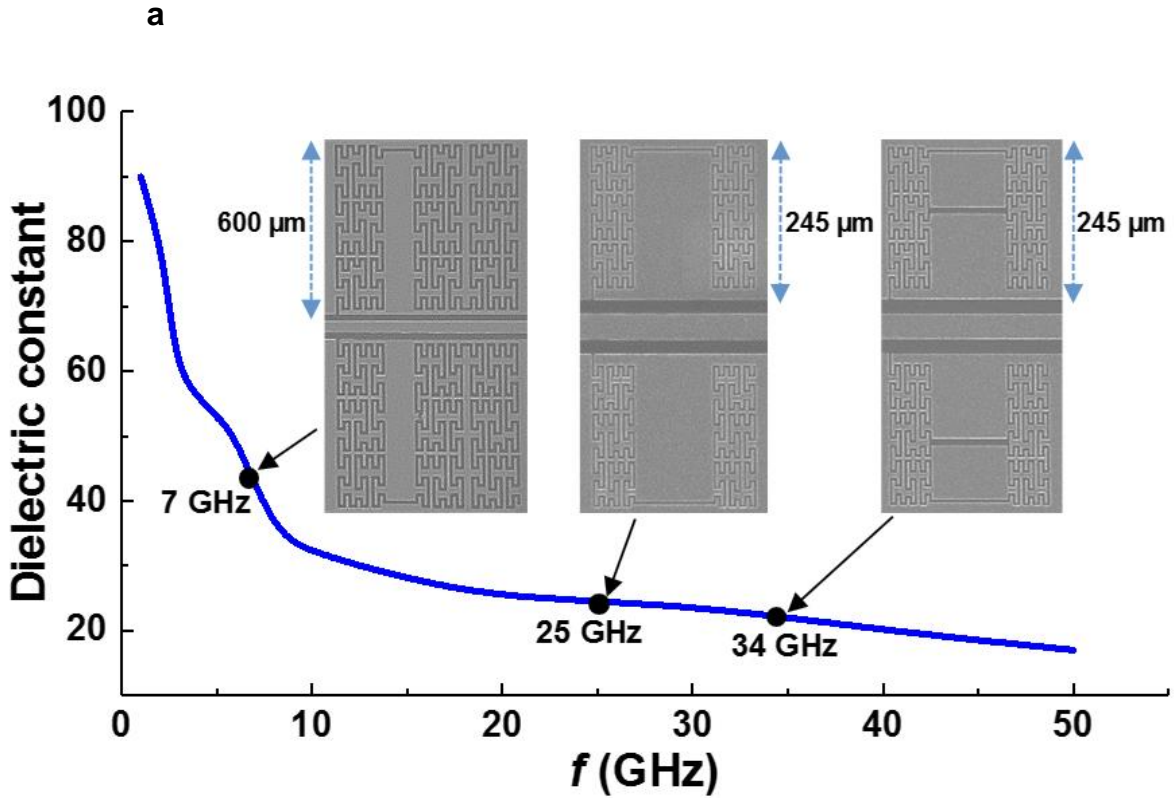

a

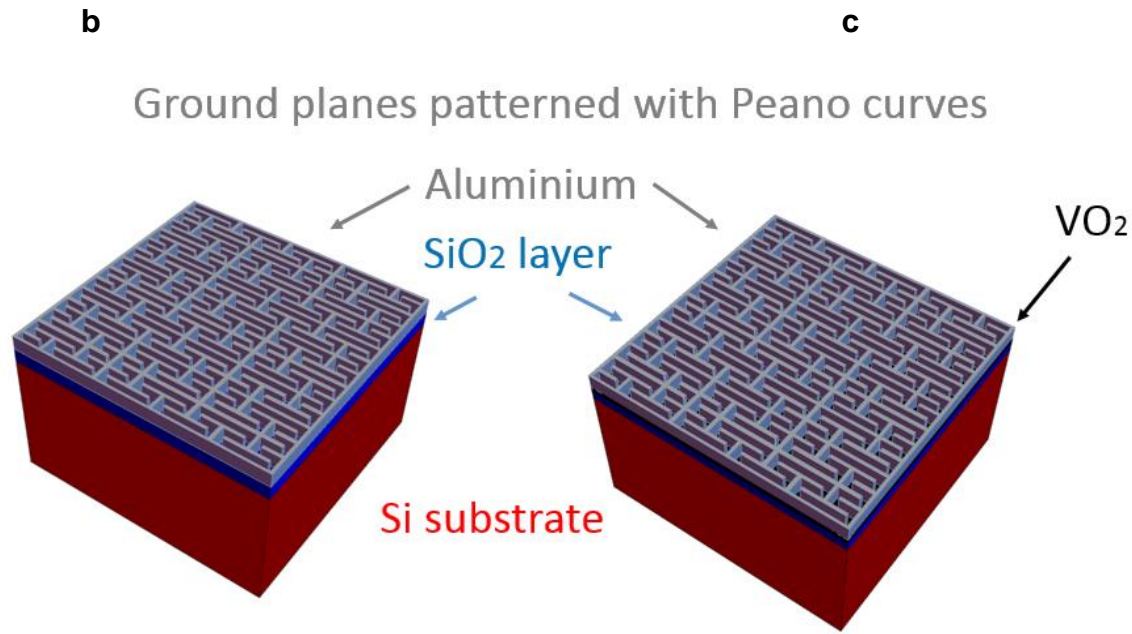

**Fig. 5| Dielectric constant extraction through frequency shift of Peano inspired defected ground plane structures at room temperature a,** Frequency dependency of the dielectric constant extracted with a variety of structures resonating at different frequencies. The same structure fabricated without and with the VO$_2$ layer below the metallization resonates at a different frequency. The frequency shift is given by the different relative effective dielectric permittivity. **b,** Substrate configuration for the measurement without the 170 nm VO$_2$ layer. **c,** Substrate configuration for the measurement with the 170 nm VO$_2$ layer just below the metallization (above the 500nm of SiO$_2$).

**Conclusions**

We have first reported new theoretical foundations for a frequency-dependent 3D Smith with 3D visualization methods for the orientation of parametric curves and used them to quantify and understand curvature reversal, while sweeping the frequency, of driving point impedances and reflection coefficients of circuits in the RF frequency bands. Further we have additionally extended the capabilities of the tool to simultaneously envision a variety of frequency dependent parameters required in the inductor design and thus proposed a unique multi-parameter display. We demonstrated by design, fabrication and measurements, original Peano reconfigurable inductors by employing the phase change VO$_2$ materials in CPW/ CMOS compatible technology on SiO$_2$. The reported inductors improve the previously reported state of art in the incipient field of VO$_2$ reconfigurable inductors design for the S, C and X band of the radio frequency spectrum. Last, a variety of first-time DGS Peano structures resonating within the mobile, satellite communications and military bands were fabricated in CPW CMOS compatible technology in order to extract based on a frequency shift, for the first time, the VO$_2$ dielectric constant within the 1 GHz-50 GHz frequency range and to propose and validate a more general and accurate model for it.

**Methods**

**Fabrication**

The devices were fabricated using a high resistivity (10000 Ω·cm) Si wafer (525 μm) as the starting substrate. A 300 nm-thick amorphous Si was deposited by low pressure chemical vapor deposition (LPCVD), to reduce the losses during measurement. A surface passivation using 500 nm sputtered SiO$_2$ was then carried out, followed by deposition of 140 nm (and 170nm)-thick VO$_2$ using a Pulsed Laser Deposition (PLD) system The film was deposited by pulsed lased deposition (PLD) using a Solmates SMP 800 system. The deposition was performed at 400 C in oxygen ambient, with a chamber pressure of 0.01 mbar. The ablated $V_2O_5$ target was placed at 60 mm distance from the wafer. Further the deposition, an annealing of 10 min at 475 C was performed without breaking the vacuum in the chamber.

The fabrication process for the Peano filters on the above substrate then commenced with a photolithography step to pattern the $VO_2$ followed dry etching to remove the $VO_2$ from the unwanted areas. A Cr-20 nm/ Al-400 nm bi-layer metal stack was then deposited by evaporation after a subsequent photolithography step on the patterned $VO_2$. This thin metallization made it possible to realize sub-micron gaps (600 nm) which is critical for extracting a good Q in the conductive state of $VO_2$ for the inductors. This was followed by deposition of a 2.4 µm-thick Al layer on these contact pads by conventional lithography followed by metal lift-off procedure to form the CPW elements with low RF losses.

**Devices simulation and characterization**

The numerical simulations of the inductor and filters were done in HFSS ANSYS commercial software relying on the finite element method (FEM) to solve Maxwell equations. Considering the full wave electromagnetic simulation technique, we used the modal solution type for the inductors and the terminal solution type for the filter simulations. The conductivity of the Al was decreased to $3.1*10^7$S/m and the $VO_2$ switches were modeled for the inductors as simple dielectrics with 20S/m losses and a loss tangent of 0 in the off state. In the on state, the $VO_2$ switches were simulated as lossy metals of a conductivity of 40,000S/m. Subsequently full inductors and filters models were built in the software according to the actual physical structure fabricated. The Peano shapes were implemented using the Equation Curve facility of the tool, their equations being written parametrically and where needed rotations and reflections were used for the very final shapes.

In the case of the filters an extensive de-embedding was used in order to extract the frequency dependent model of $VO_2$ using variable dielectric constant, tan of losses and conductivity by fitting the results obtained. In the frequency model of the relative effective dielectric permittivity all parameters available in the HFSS model were used (frequency dependency of the dielectric constant, loss tangent, conductivity losses)

The devices were measured with the Anrtisu Vector Star VNA in a Cascade Summit prober with controllable chuck temperature who was set to 20°C in the "off state" and 100°C in the on state. For the 2D graphical interpretation the measured S parameters were converted using the Anritsu Star VNA installed Microwave Office too into the desired parameters analyzed.

**3D Smith Chart implementation**

The 3D Smith Chart application is developed using the Java programming language and the following libraries and development environment are used:
- 3D rendering: OpenGL through the Java Binding for the Open GL API (JOGL2) library;
- Mathematical operations and complex data representation: the Apache Commons Mathematics Library;
- Development environment: The NetBeans IDE with Beans Binding Library for the implementation of the application GUI and JOGL2 usage.

Further implementation details about the new mode of visualization, new simulation parameters used in the paper and their 3D representation on the Riemann sphere can be found in Supplementary Section 2.

**Mathematical modelling of curvature**

The calculations for the oriented curvature was performed using Mathematica software tool by writing the frequency parametric equations of the curves analyzed.

## Acknowledgments

This work was supported by the HORIZON2020 FETOPEN PHASE-CHANGE SWITCH Project under Grant 737109. The work of E. Sanabria-Codesal was supported in part under DGCYT Grant MTM2015-64013-P.

## Author contributions
A.A.M. proposed the new quantification model of the orientation of frequency over the 3D Smith chart, use of oriented curvature in the mathematical modelling, 3D series and shunt inductor Q inductor modelling, designed, fabricated and measured the devices with VO$_2$. A.M. supervised and guided the overall 3D Smith chart tool new capabilities implementations for this article. V.A. contributed to the development of the new 3D Smith chart concepts, their actual Java implementation and revised and arranged the paper. E.S-C refined and contributed extensively to the oriented curvature mathematical computations (equations). R.A.K contributed with the fabrication and process flow in different stages, especially in obtaining the sub-micronic gaps in the Al via photolithography. A.K. contributed with the conductivity measurements of the VO$_2$. M.F-B contributed with constructive critical view in the preliminary design stages. M.C made the PLD deposition of VO$_2$. J.Z and E.C. contributed with the calibration of the measurement tool. A.S. supervised the work of A.K. A.M.I. lead the work of A.A.M and guided the implementation stages for each part of the work.

## Competing interests
The authors declare no conflict of interest.

## Materials & Correspondence
Correspondence and requests for materials should be addressed to A.A.M.

# SUPPLEMENTARY MATERIAL

## S1. On the clockwise and anticlockwise frequency dependency of Foster and non-Foster circuits on the Smith chart

### 1.1. Misleading remarks on the Counter-clockwise movement on the Smith chart

In[1] (year 2012) the authors state "The locus of the deembedded measured results rotates anti-clockwise on the Smith chart between 595MHz and 1.5GHz and this implies the presence of a non-Foster element".

In[2] (2012) "the result shows non-Foster behavior from 595MHz to 1.5GHz in the anticlockwise rotation of the traces".

In[3] (2013) "This exhibits the expected non-Foster performance by its anticlockwise rotation around the Smith chart with increasing frequency between 0.595 GHz and 1.5 GHz".

In[4] (2010) "It can be seen that both prototypes clearly show non-Foster behavior (locus of input impedance rotates counter-clockwise with frequency) within a very broad band (more than two octaves).

In[5] (1957) "The admittance function plotted by the author on the Smith chart winds counterclockwise. This is contrary to the laws of nature for passive linear networks". These laws prescribed that the plot of the admittance function wind clockwise from the low to the high frequency".

### 1.2. Counter-clockwise movements network containing Foster elements

In[6-7] ([6]Appendix B)-([7]Appendix A) the so often misleading interpretation of counter-clockwise movement on the Smith chart as a direct consequence of a non-Foster element was noticed by B. A. Munk (1929-2009) where he gives empirically examples when this happens also for Foster elements (on a limited bandwidth).

### 1.3. Correct statements on the Counter-clockwise movements network containing lossy non-Foster elements

In[8] it is stated (for the particular case of an antenna) "Usually, the impedance of any antenna moves about the Smith chart in a clockwise direction with increasing frequency. However, this behavior is not dependent on the antenna obeying the Foster reactance theorem".

### 1.4. Oriented curvature of a one port network

We will construct our proposed theory based on the notion of oriented curvature[9-10]. Given a regular plane curve $C(s)=(r(s), x(s))$, i.e. $C'(s) \neq 0$, we will call tangent vector $t(s) = C'(s)$. We consider that $C(s)$ it is parametrized by the arc length, i.e. $<t(s), t(s)>=1$. In this case, the tangent vector can be completed with a vector $n(s)=(-x'(s), r'(s))$, so that $\{t(s), n(s)\}$ form a positively oriented orthonormal base (the determinant of both vectors is positive). By using that $<t(s), t(s)>=1$, we obtain $2 <t'(s), t(s)>=0$, then $t'(s)$ and $n(s)$ have the same direction. We will call the derivative of the tangent vector $C''(s)=t'(s)$ the curvature vector of the curve. The oriented curvature $k(s)$ is the projection of this curvature vector on the normal vector, thus

$$k(s)= <n(s), C''(s)> = -x'(s)r''(s) + r'(s)x''(s) = \begin{vmatrix} r'(s) & x'(s) \\ r''(s) & x''(s) \end{vmatrix} \quad (S1)$$

The formula for the oriented curvature for an arbitrary parameterization of $C(\omega)= r(\omega)+j\, x(\omega)$ is obtained[13] by using chain rule[10]:

$$k(\omega)= \frac{-x'(\omega)r''(\omega)+r'(\omega)x''(\omega)}{(r'(\omega)^2+x'(\omega)^2)^{3/2}} = \frac{\begin{vmatrix} r'(\omega) & x'(\omega) \\ r''(\omega) & x''(\omega) \end{vmatrix}}{(r'(\omega)^2+x'(\omega)^2)^{3/2}} \quad (S2)$$

Then, the curvature sign is positive if the projection of $C''(\omega)$ on the normal vector has the same direction as $n(\omega)$ and negative in other case. As $C''(\omega)$ is the variation of the tangent vector $t(\omega)$ and the normal vector $n(\omega)$ completes the positively oriented orthonormal base $\{t(\omega), n(\omega)\}$ (i.e. it is always to the left of the tangent vector), then one gets that $k(\omega)$ is positive when the curve is turning "left" with respect to the tangent vector and negative when it turns "right". Therefore, the oriented curvature of a plane curve gives us information about the orientation of the curve and it is easy to see that the lines have zero curvature and that the curves with negative curvature are curves clockwise oriented and the curves with positive curvature are counterclockwise oriented. Consider first a one port network ended on a resistive load $r$. The input impedance $z_m(j\omega)$ is given by (S3), where $r_m(\omega)$ denotes its resistive part and $x_m(\omega)$ its reactive part, while its reflection coefficient is given by (S4). Computing the oriented curvature values for both of them we get $k_{zm}(\omega)$ (the oriented curvature of the input impedance) and $k_{\Gamma_{1zm}}(\omega)$ (the oriented curvature of the reflection coefficient) as (S5) and (S6).

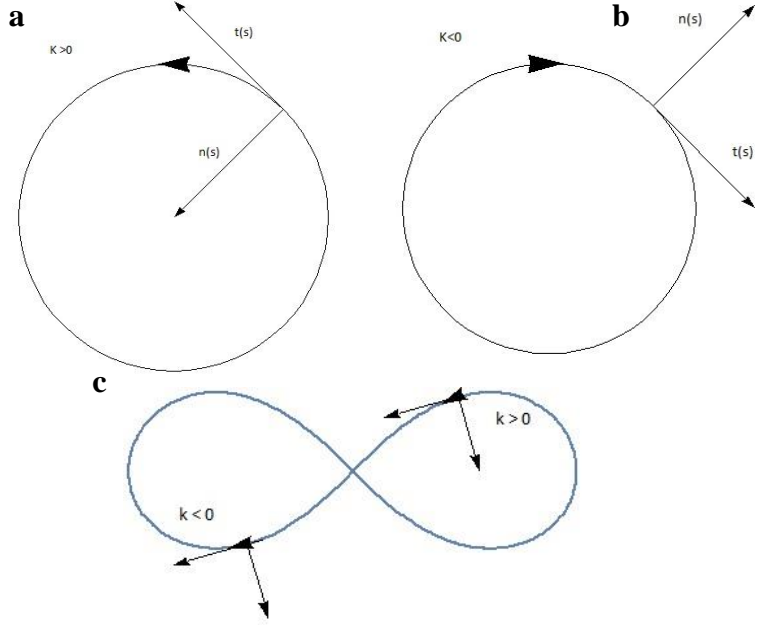

**Supplementary Fig.1: Oriented curvature of a plane curve. a,** The oriented curvature $K$ is positive. **b,** The oriented curvature K is negative **c,** The oriented curvature changes sign as one moves along it.

$$z_m(j\omega) = r_m(\omega) + jx_m(\omega) \tag{S3}$$

$$\Gamma_{1zm}(j\omega) = \frac{z_m(j\omega)/r - 1}{z_m(j\omega)/r + 1} \tag{S4}$$

$$k_{zm}(\omega) = \frac{-x_m'(\omega)r_m''(\omega) + r_m'(\omega)x_{zm}''(\omega)}{(r_m'(\omega)^2 + x_m'(\omega)^2)^{3/2}} = \frac{r_m'(\omega)^2 \left(\frac{x_m'(\omega)}{r_m'(\omega)}\right)'}{(r_m'(\omega)^2 + x_m'(\omega)^2)^{3/2}} \tag{S5}$$

$$k_{\Gamma_{1zm}}(\omega) = \frac{r_m'(\omega)^2((r + r_m(\omega))^2 + x_m(\omega)^2)\left(\frac{x_m'(\omega)}{r_m'(\omega)}\right)' + 2\,x_m(\omega)^2(r_m'(\omega)^2 + x_m'(\omega)^2)\left(\frac{r + r_m(\omega)}{x_m(\omega)}\right)'}{2\,r\,(r_m'(\omega)^2 + x_m'(\omega)^2)^{3/2}} \tag{S6}$$

Following (S 1)-(S 6) and **Supplementary Fig.1** one can easily conclude that[9-10]:

(i) If $k_{zm}(\omega) > 0$ then $z_m(j\omega)$ is counter-clockwise oriented,

(ii) if $k_{zm}(\omega) < 0$ then $z_m(j\omega)$ is clockwise oriented,

(iii) if $k_{\Gamma_{1zm}}(\omega) > 0$ then $\Gamma_{1zm}(j\omega)$ is counter-clockwise oriented

(iv) $k_{\Gamma_{1zm}}(\omega) < 0$ then $\Gamma_{1zm}(j\omega)$ is clockwise oriented.

(v) If $k_{zm}(\omega) = 0$ then $z_m(j\omega)$ is a line.

(vi) If $k_{\Gamma_{1zm}}(\omega) = 0$ then $\Gamma_{1zm}(j\omega)$ is a line.

In order that the orientation of the curves $z_m(j\omega)$ and $\Gamma_{1zm}(j\omega)$ should be the same, their curvatures must have the same sign. Paying attention to the formulas obtained for the curvature of $z_m(j\omega)$, the sign of $k_{zm}(\omega)$ depends only of $\left(\frac{x_m'(\omega)}{r_m'(\omega)}\right)'$, because the rest of the factors are squared and therefore they are always positive, including r, which is a positive constant. However, to calculate the sign of $k_{\Gamma_{1zm}}(\omega)$, we will have to take into account $\left(\frac{x_m'(\omega)}{r_m'(\omega)}\right)'$ and $\left(\frac{r+r_m(\omega)}{x_m(\omega)}\right)'$:

$$\text{If } k_{zm}(\omega) > 0 \text{ and } \left(\frac{r+r_m(\omega)}{x_m(\omega)}\right)' > 0, \text{ we obtain } k_{\Gamma_{1zm}}(\omega) > 0 \tag{S7}$$

$$\text{If } k_{zm}(\omega) < 0 \text{ and } \left(\frac{r+r_m(\omega)}{x_m(\omega)}\right)' < 0, \text{ we obtain } k_{\Gamma_{1zm}}(\omega) < 0 \tag{S8}$$

In any other case, the changes of sign between (S5) and (S6) as the angular frequency changes will depend on the sign of the numerator of the total expression (S6) and the following relation has to be obeyed so that both curvatures should have the same sign:

$$\left| r_m'(\omega)^2((r+r_m(\omega))^2 + x_m(\omega)^2)\left(\frac{x_m'(\omega)}{r_m'(\omega)}\right)' \right| > \left| 2 x_m(\omega)^2(r_m'(\omega)^2 + x_m'(\omega)^2)\left(\frac{r+r_m(\omega)}{x_m(\omega)}\right)' \right| \tag{S9}$$

This condition will assure that both curvatures have the same sign and therefore the same orientation.

## 1.5. Prove that counter-clockwise frequency dependence can occur in linear passive circuits driving point immitance

Any passive, causal, linear and stable network which can be realized by resistors, capacitors, inductors and transformers has a driving-port immitance characterized in the frequency domain by positive real functions[11-12].

Let us consider a linear one port passive network represented by the following input impedance of the (S10). Making the notations in equations (S3)-(S4) for the real and imaginary part (where $\omega$ represents angular frequency) we obtain the expressions in equations (S10)-(S14). Considering the input port impedance *r=1*) we may compute the reflection coefficient with equations (S15)-(S17).

$$z_m(s) = \frac{(s+4)(s+1)}{(s+2)(s+2.5)(s+1.5)} \tag{S10}$$

$$s = j\omega \tag{S11}$$

$$z_m(j\omega) = r_m(\omega) + jx_m(\omega) \tag{S12}$$

$$r_m(\omega) = \frac{30+27.25\omega^2+\omega^4}{56.25+48.06\omega^2+12.5\omega^4+\omega^6} \tag{S13}$$

$$x_m(\omega) = \frac{-9.5\omega-14.25\omega^3-\omega^5}{56.25+48.06\omega^2+12.5\omega^4+\omega^6} \tag{S14}$$

$$\Gamma_{1zm}(j\omega) = \frac{z_m(j\omega)-1}{z_m(j\omega)+1} = \Gamma_{zmr}(\omega) + j\Gamma_{zmi}(\omega) \tag{S15}$$

$$\Gamma_{zmr}(\omega) = -\frac{40.25+31.0625\omega^2+11.5\omega^4+\omega^6}{132.25+119.5625\omega^2+15.5\omega^4+\omega^6} \tag{S16}$$

$$\Gamma_{zmi}(\omega) = -\frac{19\omega+28.5\omega^3-2\omega^5)}{132.25+119.5625\omega^2+15.5\omega^4+\omega^6} \tag{S17}$$

Based on equations (S13) and (S14) the input impedance $z_m(j\omega)$ is plotted in **Supplementary Fig.2**. It has a clockwise orientation for *-2< ω<-0.19* and for *0.19< ω<2*, while counter-clockwise orientation for *-0.19< ω<0.19*. Using (S16) and (S17) we present in **Supplementary Fig.2.b** the Smith chart plot of the reflection coefficient $\Gamma_{1zm}(j\omega)$. It has a clockwise orientation for *-2< ω<-0.28* and for *0.28< ω<2*, while counter-clockwise orientation for *-0.28< ω<0.28*. Using (S5) and (S6) we plot the oriented curvatures of $z_m(j\omega)$ and $\Gamma_{1zm}(j\omega)$ in **Supplementary Fig.2. c**: $k_{zm}(\omega) < 0$ for *-2< ω<-0.19* and for *0.19< ω<2*, while $k_{zm}(\omega) > 0$ for *-0.19< ω<0.19*.

For $k_{\Gamma_{1zm}}(\omega)$ we get $k_{\Gamma_{1zm}}(\omega) < 0$ for -2< ω<-0.28 and for 0.28< ω<2, while $k_{\Gamma_{1zm}}(\omega) > 0$ -0.28< ω<0.28. As described in (i) and (ii) the changes of sign in the curvature correspond to the changes of orientation in the $z_m(j\omega)$ and $\Gamma_{1zm}(j\omega)$. In **Supplementary Fig.2. d** one can see $x_m(\omega)$, it is a strictly monotonic decreasing function while sweeping angular frequency for -2<ω<-2 since its derivative $\frac{dx_m(\omega)}{d\omega} < 0$. It worth noting that the curvature changes in $z_m(j\omega)$ and $\Gamma_{1zm}(j\omega)$ occur for different values of ω and that counter-clockwise motion occurs for both of them on a limited frequency range.

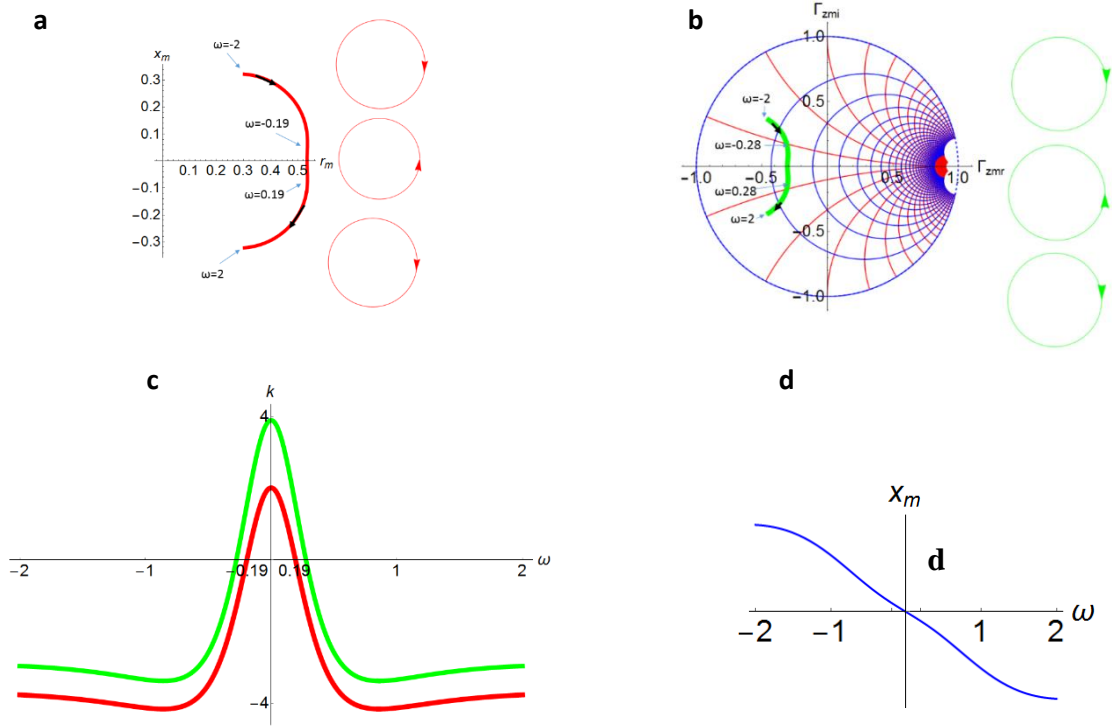

**Supplementary Fig. 2: Orientation and oriented curvature sign changes for the input impedance $z_m(j\omega)$ in (S 10) and for its corresponding reflection coefficient $\Gamma_{1zm}(j\omega)$. a,** The input impedance $z_m(j\omega)$ (red), has mixed orientations, with orientation changes corresponding to the zeros of its oriented curvature $k_{zm}(\omega)$. **b,** The reflection coefficient $\Gamma_{1zm}(j\omega)$ (green) has also mixed orientations on the Smith chart with the changes in it corresponding to the zeros of its oriented curvature $k_{\Gamma_{1zm}}(\omega)$. **c,** The oriented curvatures for $z_m(j\omega)$ and $\Gamma_{1zm}(j\omega)$: $k_{zm}(\omega)$ (red) and $k_{\Gamma_{1zm}}(\omega)$ (green). They change sign at ω= ±0.19 and ω= ±0.28. **d,** The $x_m(\omega)$ has no monotony changes for -0.28< ω<0.28.

Circuits including non-Foster elements can surely exhibit too counter-clockwise Smith chart rotations on certain frequency ranges. The non-Foster elements, which do not obey the Foster reactance theorem[13] (which is for purely reactive networks), given in (S18) (where *X* stands for reactance and *B* stands for the susceptance) play a major role in the bandwidth enhancement, being extensively used in small antenna designs. However, in practical realization they are always lossy[13], (being not purely reactive). In **Supplementary Fig. 3.** we can see the $S_{11}$ parameters of a network including Non-Foster elements (provided with the extremely kind amiability of Prof. R.W. Ziolkowski), employed in [13].

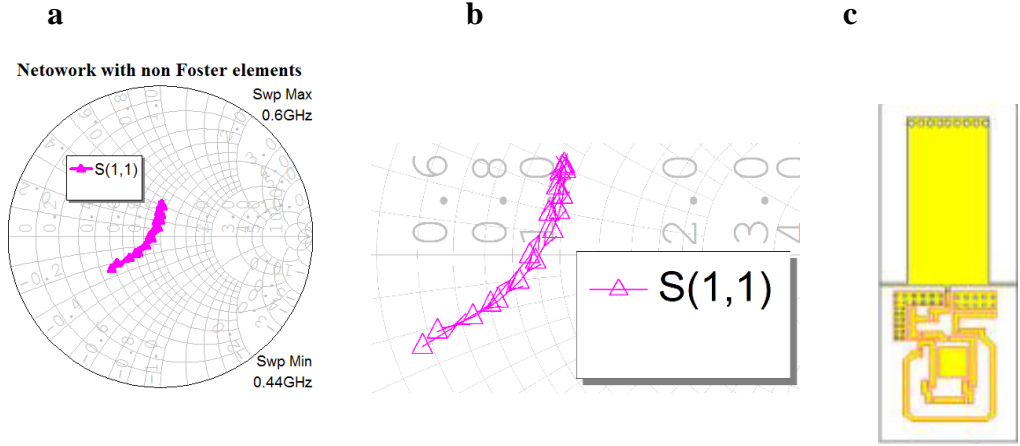

**Supplementary Fig. 3: Circuit with non Foster elements**[13] (S parameters provided with kindness by R.W. Ziolkowski). Clockwise and counter-clockwise movement of the S parameters occurs but may be overlooked and may be uneasy to be detected on a classical Smith chart. Please look on **Supplementary Video**. The circuit changes orientation of the $S_{11}$ parameter at 0.49 GHz. Circuit is analyzed between 0.44 GHz and 0.60 GHz.

### 1.6. Foster circuits –purely clockwise movement

Let us first consider a simple one port circuit (**Supplementary Fig. 4a**) composed just by ideal lossless inductors and capacitors: its driving port impedance reactance $x_F(\omega)$ is given by (S19) (a), where which is monotonically increasing with the angular frequency $\omega$ - Foster theorem[7-8,12,14-15] (S18). Its 1-port $\Gamma_{1F}(j\omega)$ reflection coefficient is given in (S19) (b). Considering now the same Foster network in a two port (**Supplementary Fig. 4b**) connection (assuming identical ports impedances $r$) one can compute the reflection coefficient as $\Gamma_{2F}(j\omega)$ (S19) (c). Calculating now the associated oriented curvatures of (S19) (a)-(c), we obtain $k_z(\omega)$, $k_{\Gamma_{1F}}(\omega)$, $k_{\Gamma_{2F}}(\omega)$, i. e. the expressions (S20) (a), (b) and (c). The oriented curvature of the one port Foster network $k_{\Gamma_{1F}}(\omega)$ = -1 implies that the reflection coefficient of the 1-port Foster networks will have a clockwise orientation as $\omega$ increases (see statement (ii)). The oriented curvature the $k_{\Gamma_{2F}}(\omega)$ = -2 implies the same (since its sign is negative).

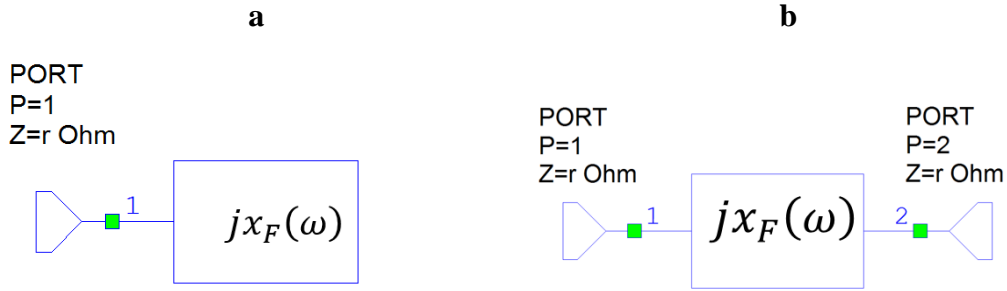

**Supplementary Fig. 4: One port and two port networks with Foster circuit elements**. **a,** One port circuit containing only Foster elements $z_m(j\omega) = jx_F(\omega)$  **b,** Two port circuit containing only Foster elements.

$$\frac{dx_F(\omega)}{d\omega} > 0 \text{ and } \frac{dB_F(\omega)}{d\omega} > 0 \quad (S18)$$

(a) $z_F(j\omega) = jx_F(\omega)$   (b) $\Gamma_{1F}(j\omega) = \frac{jx_F(\omega)/r - 1}{jx_F(\omega)/r + 1}$   (c) $\Gamma_{2F}(j\omega) = \frac{jx_F(\omega)}{jx_F(\omega) + 2}$   (S19)

(a) $k_{Fz}(\omega) = 0$, (b) $k_{\Gamma_{1F}}(\omega) = -\dfrac{rx_F'(\omega)}{(r^2 + x_F(\omega)^2)\sqrt{\dfrac{r^2 x_F'(\omega)^2}{(r^2 + x_F(\omega)^2)^2}}} = -1$,  (c) $k_{\Gamma_{2F}}(\omega) = = -\dfrac{2(4 + x_F(\omega)^2)\sqrt{\dfrac{x_F'(\omega)^2}{(4 + x_F(\omega)^2)^2}}}{x_F'(\omega)} = -2$   (S20)

The reflection coefficients $\Gamma_{1F}(j\omega)$ and $\Gamma_{2F}(j\omega)$ are direct inversive transformation of $z_F(j\omega)$ (simple Möbius transformation)[16,17] having the form (S21) (where $a, b, c, d$ are constants and $f(\omega)$ any real valued function)

$$\Gamma(j\omega) = \frac{a*jf(\omega)+b}{c*jf(\omega)+d} \tag{S21}$$

and thus[17] they map the oriented line of $jx_F(\omega)$ (imaginary axes of the impedance plane) into circles with clockwise orientation, i. e. the curvature $k_{\Gamma 1F}(\omega) < 0$ and $k_{\Gamma 2F}(\omega) < 0$. Additionally, $1/|k_{\Gamma 1F}(\omega)| = 1$, $1/|k_{\Gamma 2F}(\omega)| = 0.5$, thus the reflection coefficients will move on the unit circle, respectively on a 0.5 radius circle.

### 1.7. Non-Foster circuits –purely anti-clockwise movement

Let us first consider a simple one port circuit (**Supplementary Fig. 5a**) composed just by ideal lossless inductors and capacitors but with negative inductances and capacitances[14-15]: the driving point reactance $x_{NF}(\omega)$ (susceptance) obey (S22) while it can be always written as (S 23) where $x_{F1}(\omega)$ is a Foster reactance obeying (S18). Its 1-port $\Gamma_{1NF}(j\omega)$ reflection coefficient is given in (S 25). Considering now the same Non-Foster network in a two port (**Supplementary Fig. 5 b**) connection (assuming identical ports impedances $r$) one can compute the reflection coefficient as $\Gamma_{2NF}(j\omega)$ (S24) (c). Calculating now the associated oriented curvatures of (S24) (a)-(c), i. e. $k_{Nz}(\omega), k_{\Gamma 1NF}(\omega), k_{\Gamma 2NF}(\omega)$, we get (S25) (a), (b) and (c).The oriented curvature of the one port Foster network $k_{\Gamma 1NF}(\omega)= 1$ implies (statement (iii)) that the reflection coefficient of the non-Foster 1 port networks will have a counter-clockwise orientation as $\omega$ increases. The oriented curvature the $k_{\Gamma 2F}(\omega)= 2$ implies the same for the two port case scenario ( since its sign is positive).

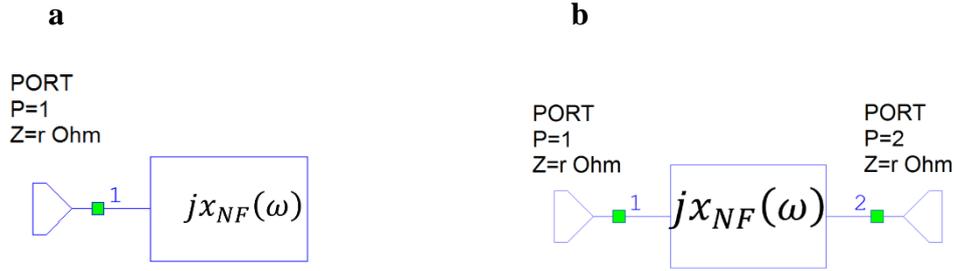

**Supplementary Fig. 5: One port and two port networks with purely non-Foster circuit elements**[14-15]. **a,** One port circuit containing only Non-Foster elements $z_m(j\omega) = jx_{NF}(\omega)$ **b,** Two port circuit containing only Non- Foster elements.

$$\frac{dx_{NF}(\omega)}{d\omega} < 0 \text{ and } \frac{dB_{NF}(\omega)}{d\omega} < 0 \tag{S22}$$

$$x_{NF}(\omega) = -x_{F1}(\omega) \tag{S23}$$

(a) $z_{NF}(j\omega) = -jx_{F1}(\omega)$ (b) $\Gamma_{1NF}(j\omega) = \frac{-jx_{F1}(\omega)/r-1}{-jx_{F1}(\omega)/r+1}$ (c) $\Gamma_{2NF}(j\omega) = \frac{-jx_F(\omega)}{-jx_F(\omega)+2}$ (S24)

(a) $k_{NFz}(\omega)= 0$, (b) $k_{\Gamma 1NF}(\omega)=\frac{rx_F'(\omega)}{(r^2+x_F(\omega)^2)\sqrt{\frac{r^2 x_F'(\omega)^2}{(r^2+x_F(\omega)^2)^2}}} = 1$, (c) $k_{\Gamma 2NF}(\omega)=\frac{2(4+x_F(\omega)^2)\sqrt{\frac{x_F'(\omega)^2}{(4+x_F(\omega)^2)^2}}}{x_F'(\omega)} = 2$ (S25)

The reflection coefficients $\Gamma_{1NF}(j\omega)$ and $\Gamma_{2NF}(j\omega)$ are indirect inversive transformation of $jx_{F1}(\omega)$ (simple Möbius transformation)[16,17,18] having the form (S 26) (where $a, b, c, d$ are constants and $f(\omega)$ any real valued function)

$$\Gamma(j\omega) = \frac{a*conj(jx_{F1}(\omega))+b}{c*conj(jx_{F1}(\omega))+d} = \frac{-ajx_{F1}(\omega)+b}{-cjx_{F1}(\omega)+d} \tag{S26}$$

and thus[17] they map the oriented line of $jx_F(\omega)$ (imaginary axes of the impedance plane) into circles with counter-clockwise orientation, i. e. $k_{\Gamma 1F}(\omega) > 0$ and $k_{\Gamma 2F}(\omega) > 0$. Additionally, $1/|k_{\Gamma 1F}(\omega)| = 1$, $1/|k_{\Gamma 2F}(\omega)| = 0.5$, thus the reflection coefficients will move on the unit circle, respectively on a 0.5 radius circle.

## S2. New frequency orientation quantification, series and shunt inductance representations and inductor frequency dependent quality factors implementations over the 3D Smith chart tool

### 2.1. 3D Smith chart concept previous capabilities

The 3D Smith chart tool was released in 2017 and is based on the articles[18-22]. The 3D Smith chart tool generalizes the Smith chart[23]. The Smith chart became throughout the years a tool widely used in engineering in different areas for displaying of reflection coefficients[24-31] being extensively used in the design stage or measurement phase of a variety of circuits in the RF, microwaves and THz frequency range.

The Smith chart main equation is given in (S27) and is applied to the grid of normalized impedance plane resulting in **Supplementary Fig. 6 a**. Equation (S27) maps the right half plane of the impedance plane into the unit circle of the reflection coefficients plane. The left half plane is mapped in its exterior **Supplementary Fig. 6 b** which extends to infinity. For this reason, circuits exhibiting negative part of the input impedance cannot be analyzed on the Smith chart, while with difficulty on the extended Smith chart.

$$\Gamma(z) = \frac{z-1}{z+1} \quad z \in \{\mathbb{C}\} \tag{S27}$$

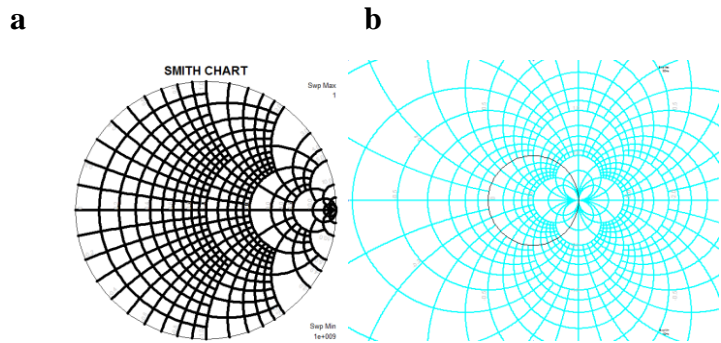

**Supplementary Fig. 6: Smith chart and extended 2D Smith chart**. **a,** The regions with reflection coefficient bigger than unity in absolute value cannot be seen  **b,** The regions with reflection coefficient bigger than unity extend to infinity thus the chart cannot be easily used for circuits with negative part of the real part of the input impedance.

The 3D Smith chart main equation is also (S27) but it is considered that $z \in \mathbb{C} \cup \{\infty\}$ and using Riemann- Maxime Bochert representation of Mobius transformations[16-17] ( in order to preserve circle shapes in a compact space) we get based on **Supplementary Fig. 7 a** and **b** in **Supplementary Fig. 8**.

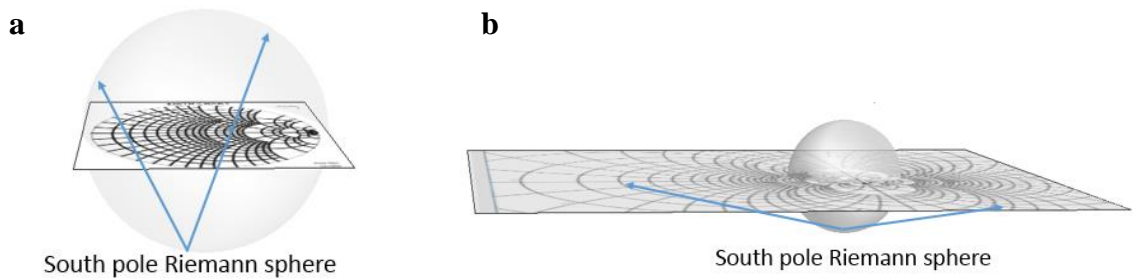

**Supplementary Fig. 7: Smith chart and extended 2D Smith chart mapping on the Riemann sphere**. **a,** The regions with reflection coefficients magnitude smaller than unity are mapped into the north hemisphere. **b,** The regions with reflection coefficient magnitudes bigger than unity are mapped into the South hemisphere.

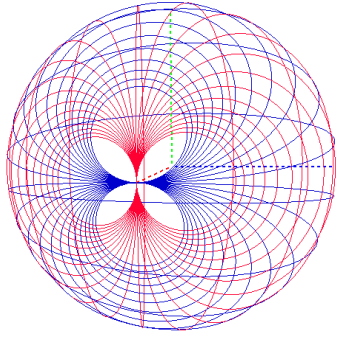

**Supplementary Fig. 8: 3D Smith chart:** South hemisphere: circuits with reflection coefficients magnitude bigger than unity. North hemisphere: circuits with reflection coefficients magnitude smaller than unity. East: Inductive, West: Capacitive.

## 2.2. Implementations details for the 3D Smith Chart Application

The 3D Smith Chart application is developed using the Java programming language and the following libraries are used:
- For the 3D rendering the OpenGL API is being used through the Java Binding for the Open GL API (JOGL2) library with multi-platform support on 32/64 bits;
- The Apache Commons Mathematics Library is being used for mathematical computations and program data structures (representation of complex numbers, fast matrix multiplication representation);

The NetBeans IDE with Beans Binding Library is used to build the GUI of the application.

### 2.2.1. Frequency visualization

A new mode of visualization of the frequency associated to each point of the 3D space curve (computed using the $S_{11}$ (complex) reflection parameter) has been developed for the present paper. Each frequency that corresponds to a point of the 3D curve will be displayed as a segment on the line that passes from the center of the 3D sphere and the point of the 3D curve. The length of the segment will be given by the normalized frequency and the direction will be outwards of the surface of the 3D sphere.

To visualize the frequency dependency, the following steps are done:

For each point $P$ of the 3D curve:

Compute the direction vector defined by the center of the sphere and the point of the 3D curve

$$dir(X,Y,Z) = \big(P(X) - sphereCenter(X), P(Y) - sphereCenter(Y), P(Z) - sphereCenter(Z)\big)$$

Normalize the direction vector

$$dir_N(X,Y,Z) = \frac{dir(X,Y,Z)}{|dir(X,Y,Z)|} \tag{S28}$$

Normalize the frequency associated to each point of the 3D curve

$$f_N = \frac{f}{Max(f)} \tag{S29}$$

Compute $P_{END}$

$$P_{END}(X,Y,Z) = (dir_N(X)*f_N, dir_N(Y)*f_N, dir_N(Z)*f_N) \tag{S30}$$

Draw the segment $[P_{START} P_{END}]$ where $P_{START}$ is P, the point of the 3D curve.

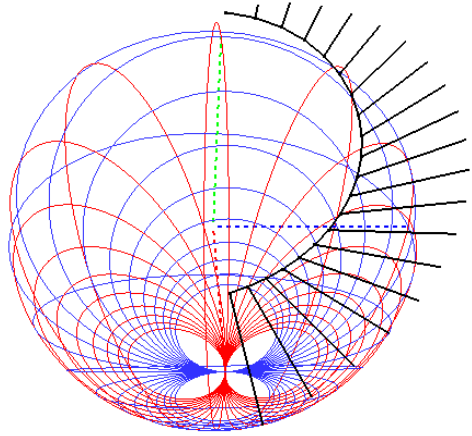

**Supplementary Fig. 9**: Frequency displayed as segment on a LSeries 3D curve.

**Note**: if multiple circuits (multiple touchstone files) are represented then $Max\,(f)$, used in the normalization of the frequency to obtain $f_N$, is the one greater across all the circuits used.

### 2.2.2. New parameters implemented

For the present paper, the following new parameters have been implemented in the 3D Smith Chart application:
- LSeries
- LShunt
- QualityFactor for LSeries/LShunt

**LSeries**

LSeries is computed using the following formula:

$$LSeries(\omega) = Imag\left(\frac{R(1+S_{11}-S_{12}S_{21}+S_{22}+S_{11}S_{22})}{2*Pi*f*2*S_{21}}\right) \tag{S31}$$

Then $LSeries(\omega)$ is normalized

$$LSeries_N(\omega) = \frac{LSeries(\omega)}{Max\,(LSeries(\omega))} \tag{S32}$$

To obtain the 3D representation of LSeries the following transformation is done

$$LSeries_{3d}(\omega) = (LSeries_N(\omega) + 1) * S_{11_{3d}}(j\omega) \tag{S33}$$

Where $S_{11_{3d}}(j\omega)$ is[21]

$$S_{11_{3d}}(j\omega) = Imag\left(\frac{2a(\omega)}{1+|S_{11}(\omega)|^2}, \frac{2b(\omega)}{1+|S_{11}(\omega)|^2}, \frac{1-|S_{11}(\omega)|^2}{1+|S_{11}(\omega)|^2}\right) \tag{S34}$$

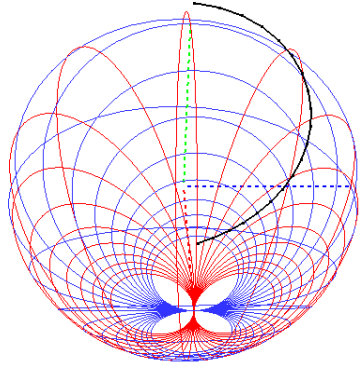

**Supplementary Fig. 10**: LSeries 3D curve (black) displayed above the 3D Smith chart unit ball. Please note that unlike the S parameters which are on the unit sphere, the normalized inductance in the 3D space.

**LShunt**

LShunt is computed using the following formula:

$$LShunt(\omega) = Imag\left(\frac{1}{y_{11}}\right) * \frac{1}{2*PI*f} \quad (S\ 35)$$

Where $y_{11}$ is

$$y_{11} = \left(\frac{1+S_{12}S_{21}+S_{22}-S_{11}(1+S_{22})}{R(1+S_{11}-S_{12}S_{21}+S_{22}+S_{11}S_{22})}\right) \quad (S\ 36)$$

Then $LShunt(\omega)$ is normalized

$$LShunt_N(\omega) = \frac{LShunt(\omega)}{Max\ (LShunt(\omega))} \quad (S37)$$

To obtain the 3D representation of LShunt the following transformation is done

$$LShunt_{3d}(\omega) = (LShunt_N(\omega) + 1) * S_{11_{3d}}(j\omega) \quad (S38)$$

Where $S_{11_{3d}}(j\omega)$ is defined in (S 34).

**Note**: if multiple circuits (multiple touchstone files) are represented then the $Max\ (LSeries(\omega))$ / $Max\ (LShunt(\omega))$, used in the normalization step to obtain $LSeries_N$ / $LShunt_N$, is the one greater across all the circuits used.

**QualityFactor for LSeries/LShunt**

The QualityFactor is computed using the following formula:

$$Q(\omega) = -\frac{Imag(y_{11})}{Real(y_{11})} \quad (S39)$$

The $Q(\omega)$ is normalized

$$Q_N(\omega) = \frac{Q(\omega)}{Max\ (Q(\omega))} \quad (S40)$$

**Note**: if multiple circuits (multiple touchstone files) are represented then the $Max\ (Q(\omega))$, used in the normalization step to obtain $Q_N(\omega)$, is the one greater across all the circuits used.

Using the 3D representation of the LSeries / LShunt curve we can use the normal plane of the curve to associate to each point of the curve its quality factor as a cylinder of variable radius. This process will be described in the next section.

### 2.2.3. LSeries / LShunt representation in 3D using the QualityFactor

For the 3D representation of the LSeries / LShunt, the normalized LSeries / LShunt value and the $S_{11}$ transmission parameter are used and thus each point of the 3D curve is computed using the following formula:

$$L_{3d}(\omega) = (L_N(\omega) + 1) * S_{11_{3d}}(j\omega) \qquad (S41)$$

For the 3D representation that uses the quality factor as a 3D variable radius cylinder, the following steps have been implemented for each segment of the 3D space curve:

### a) Draw the 3D cylinder

For each segment $[P_i P_{i+1}]$ of the LSeries / LShunt 3D space curve, a cylinder is drawn along the Z axis with the base of the cylinder centered in $P_i$ and the top of the cylinder centered in $P_{i+1}$. The radiuses of the base and of the top of the cylinder will have the values associated to the normalized quality factors for point $P_i$ and $P_{i+1}$ respectively. The height of the cylinder will have the value of the length of the $[P_i P_{i+1}]$ segment of the curve.

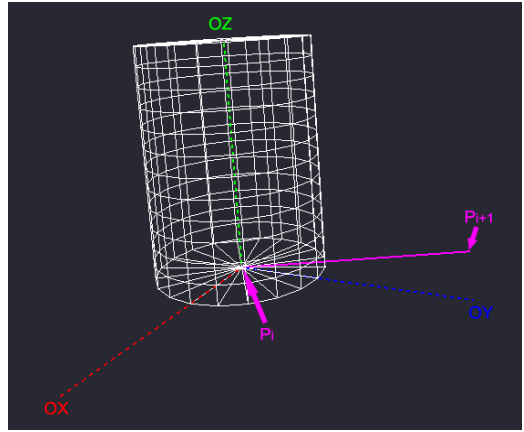

**Supplementary Fig. 11**: 3D cylinder for the $[P_i P_{i+1}]$ segment of the curve drawn along the Z axis.

### b) Compute transformation parameters for the 3D cylinder

To align the 3D cylinder along the $[P_i P_{i+1}]$ segment the following parameters are computed:
- the angle between the OZ axis and the $[P_i P_{i+1}]$ segment as the dot product
- the axis around which the 3D cylinder will be rotated as the cross product between the OZ axis and the $[P_i P_{i+1}]$ segment

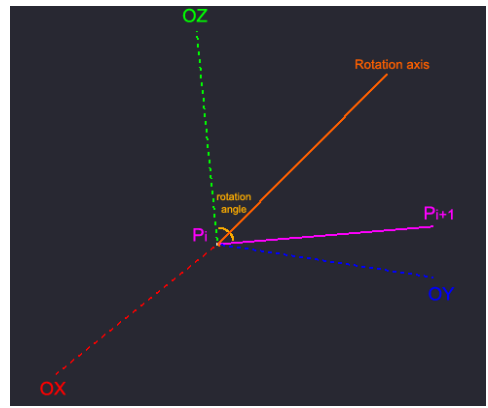

**Supplementary Fig. 12**: Transformation parameters (rotation angle and rotation axis) for the 3D cylinder.

### c) Apply the transformation to the 3D cylinder

The 3D cylinder will be aligned along the $[P_i P_{i+1}]$ segment ($P_i$ will be at the center of the base of the cylinder and $P_{i+1}$ will be at the center of the top of the cylinder) by rotating the 3D cylinder around the rotation axis and using the angle computed in the previous step.

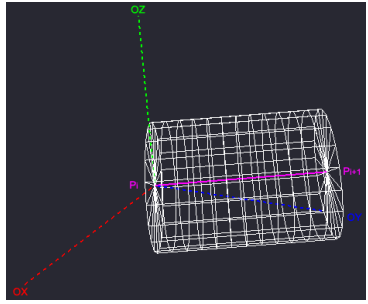

**Supplementary Fig. 13**: The 3D cylinder transformed so that it wraps around the [P$_i$P$_{i+1}$] segment of the curve.

The steps described above will be applied to each segment of the 3D space curve and the result can be seen in the picture **Supplementary Fig. 14** below:

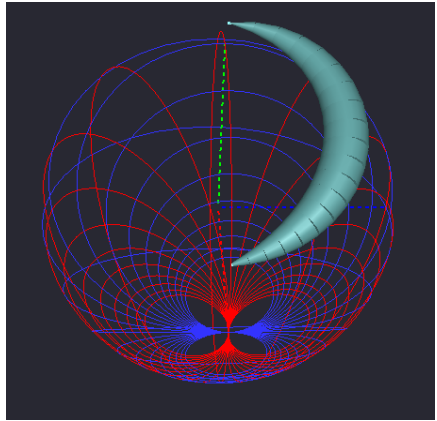

**Supplementary Fig. 14**: Final representation using 3D cylinders in solid and lit mode.

## S3. Inductances and quality factors of inductors on the 3D Smith chart

It is important to stress that there are several models for extracting the inductance of a high frequency range inductor. For example the models in[34-35] use the Y$_{21}$ to extract its value (or equivalently this means using the parameter B of the ABCD matrix):

$$\begin{bmatrix} V_1 \\ I_1 \end{bmatrix} = \begin{bmatrix} A & B \\ C & D \end{bmatrix} \bullet \begin{bmatrix} V_2 \\ -I_2 \end{bmatrix} \quad (S42)$$

$$\begin{bmatrix} A & B \\ C & D \end{bmatrix} = \begin{bmatrix} 1 & j\omega L \\ 0 & 1 \end{bmatrix} \quad (S43)$$

Using the simplified PI model of an inductor shown in **Supplementary** Fig. 15 we may see that this model neglects in its implementation the shunt capacitance:

$$j\omega L_{series} = B = (R_s + j\omega L) || (\frac{1}{j\omega C_s}) \quad (S44)$$

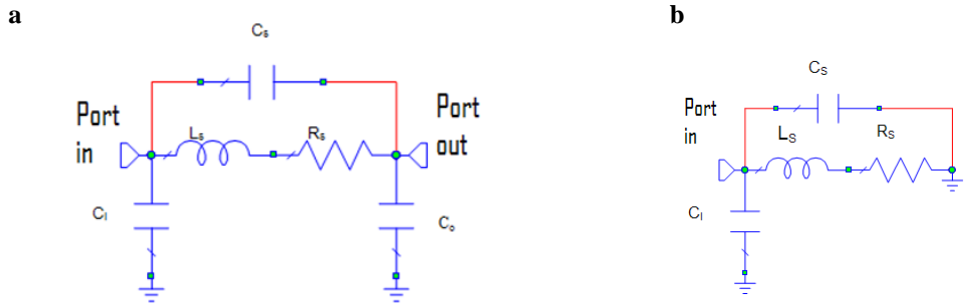

**Supplementary Fig. 15**: Simplified Pi model of an inductor. **a,** simplified Pi model **b**, Simplified Pi model with a port connected to the ground.

In[36-38] authors use the $Y_{11}$ (or equivalently parameter C) model for computing thus, in their model the second port is connected to the ground (**Supplementary Fig.15 b**). For the simplified model in **Supplementary Fig. 15** this makes $L_{shunt}$ be extracted from (S45)

$$j\omega L_{shunt} = (R_s + j\omega L) || (\frac{1}{j\omega C_s}) || (\frac{1}{j\omega C_1}) \qquad (S45)$$

Neverthless both models can be alternatiely used, since an inductor can be placed in between various circuits in various configurations[39]. Concerning the quality factor we adopt the common used model (S39)[34-38]. In **Supplementary Fig. 16** we can see the performances of the newly designed Peano inductor in respect to the previously reported inductor with $VO_2$ in the same technology ( in the conductive, on state of $VO_2$).

Considering the new designed inductors extracted inductances and quality factor for two different frequency ranges: consider **Supplementary Fig. 17 - Supplementary Fig. 21.**

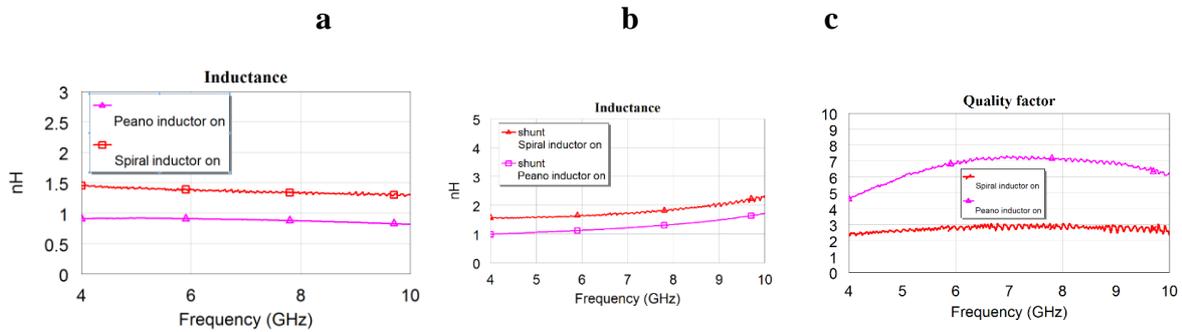

**Supplementary Fig.16**: Comparative analyze of the new Peano inductor in the conductive state of $VO_2$ versus previously reported reconfigurable inductor based on $VO_2$ in the on state. (4GHz-10 GHz): a series model of inductance, b: shunt model, c: quality factor. Peano inductor (purple), previous[34] reported inductor with $VO_2$ –red.

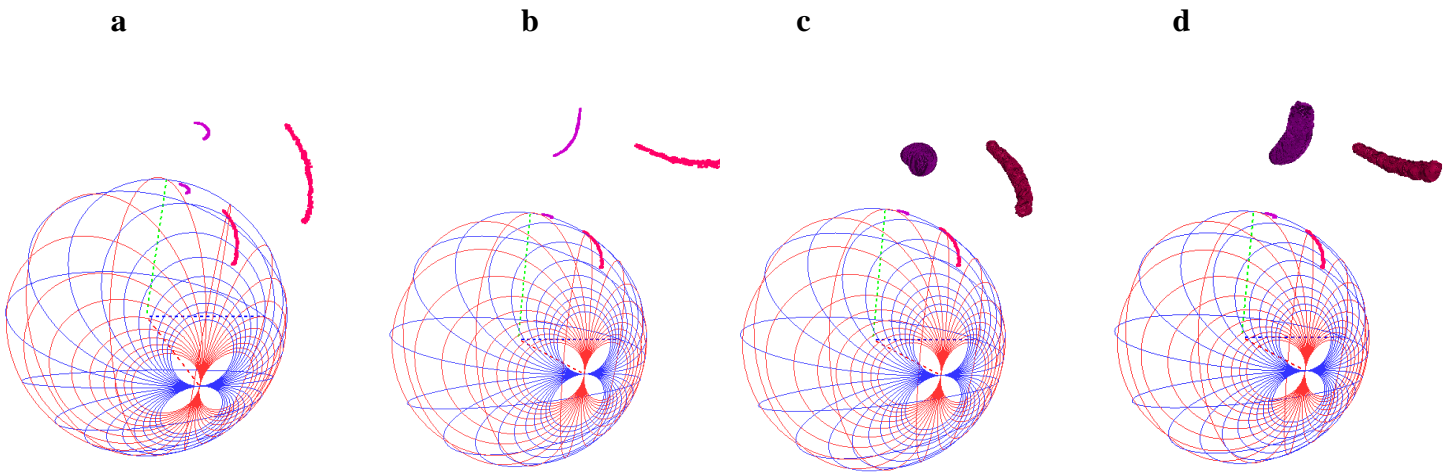

**Supplementary Fig. 17**: Comparative analyze of the new Peano inductor in the conductive state of VO$_2$ versus previously reported[34] reconfigurable inductor based on VO$_2$ in the on state. (4GHz-10 GHz) on the 3D Smith chart (corresponding to Supplementary Fig. 21 Fig): a series model of inductance, b: shunt model, c: quality factor. On the 3D smith chart one can see the S$_{11}$ parameter to which they correspond. Please check video for better visualizations.

**Supplementary Fig. 17 a** contains more (indirect) information about the shunt inductance too, it shows that up to 10 GHz both inductors are still in the East hemisphere. In **Supplementary Fig. 17 b** we can see how the shunt inductance model start being non-linear. In **Supplementary Fig. 17 c** and **d** we can see both quality factors plotted along the inductances, their frequency dependency being visible. In **Supplementary Fig. 18** we can see the clockwise orientation with increasing frequency of the S$_{11}$ reflection parameter for both inductors.

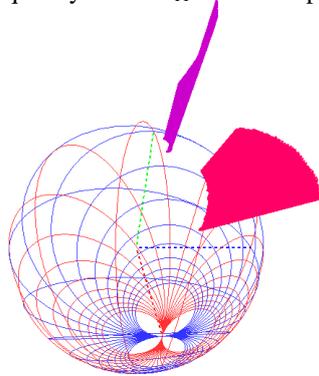

**Supplementary Fig. 18**: Comparative analyze of the new Peano inductor in the conductive state of VO$_2$ versus previously reported reconfigurable inductor based on VO$_2$ in the on state. (4GHz-10 GHz): up to 10 GHz, both inductors stay in the inductive hemisphere (East), rotating clockwise as frequency increases.

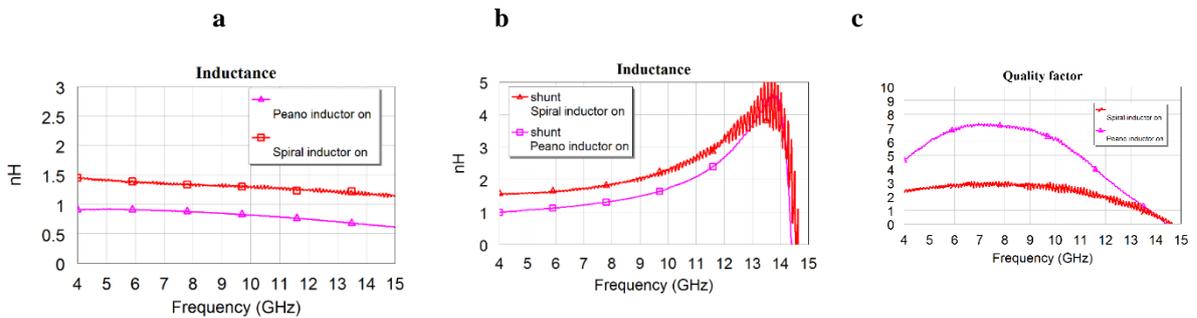

**Supplementary Fig. 19**: Comparative analyze of the new Peano inductor in the conductive state of VO$_2$ versus previously reported reconfigurable inductor based on VO$_2$ in the on state. (4GHz-15 GHz): **a**, series inductance, b: shunt inductance, c: quality factor.

a          b          c          d

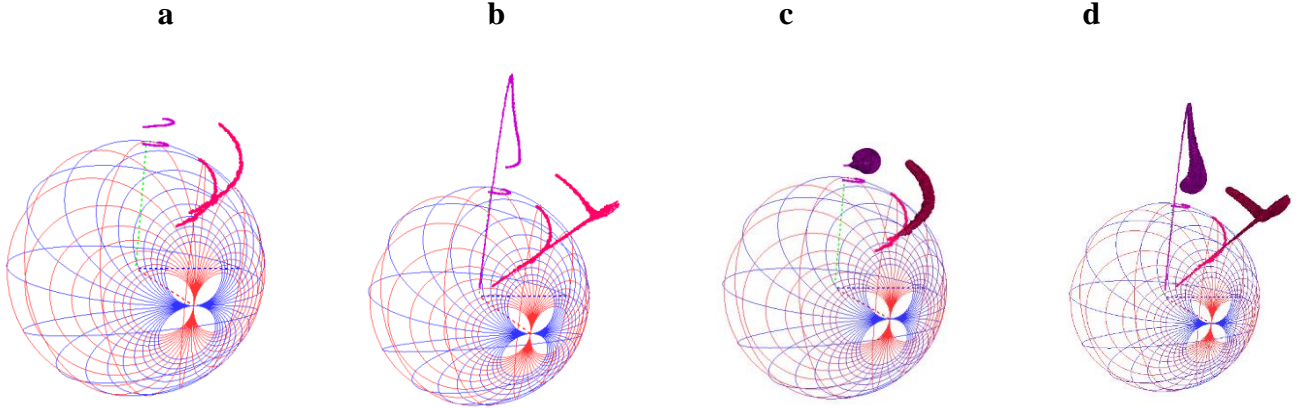

**Supplementary Fig. 20**: Comparative analyze of the new Peano inductor in the conductive state of VO$_2$ versus previously reported reconfigurable inductor based on VO$_2$ in the on state. (4GHz-15 GHz) on the 3D Smith chart: a, series inductance, b: shunt inductance, c: quality factor along the series inductance, d: quality factor along the shunt inductance. On the 3D smith chart one can see the S$_{11}$ parameter to which they correspond.

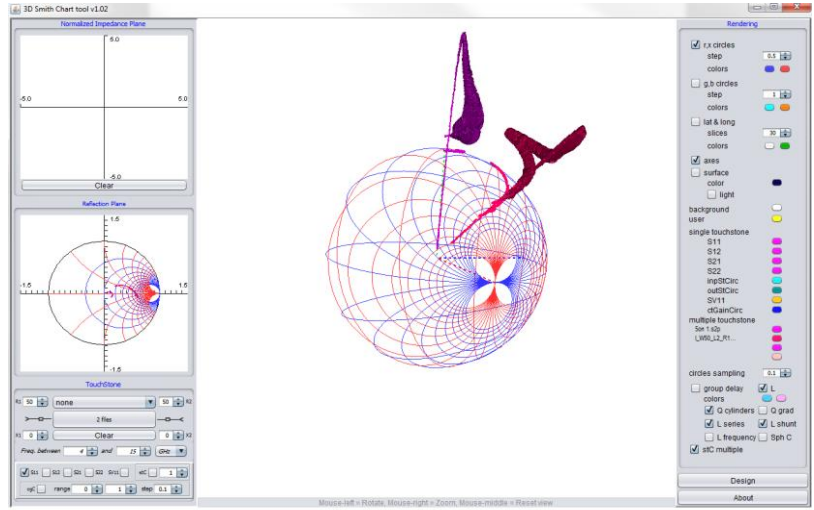

**Supplementary Fig. 21**: 3D Smith chart tool visualization of series and shunt inductances together with the quality factor on the 4-15 GHz frequency range.

Considering the simplified model in **Supplementary Fig.15** it is important to stress that if R$_S$=0 then the zeros of of the imaginary part of Y$_{11}$ are automatically zeros ( on the angular frequency axis) of the S$_{11}$ parameter, since by using elementary ABCD matrices and one can easy determine the zeros of the im (Y$_{11}$) as:

$$\omega_{imy11} = \frac{1}{\sqrt{L_S * C_i + L_S * C_S}} \tag{S 46}$$

While the zeros (on the angular frequency axis) of the imaginary part of S$_{11}$ include the zeros of the Y$_{11}$

$$\omega_{imS11} = \frac{1}{\sqrt{L_S * C_i + L_S * C_S}} \, , \, \frac{\sqrt{-L_S + 2 * C_i * Z}}{\sqrt{L_S * C_i + L_S * C_S}} \tag{S 47}$$

**Our Peano inductors:**

A zoom of the VO$_2$ switch of the fabricated inductor is shown in **Supplementary Fig. 22**. It is important to stress that our aim was to minimize resistive losses in the on state, thus we wished that the VO$_2$ switch (gap in the Al at the top, where VO$_2$ beneath is touched) should be as small in width as possible (while maintaining classical CMOS technology using standard microelectronic processes). This was desired since the conductivity of the VO$_2$ in the on state is limited on SiO$_2$/Si substrates.

**a,**

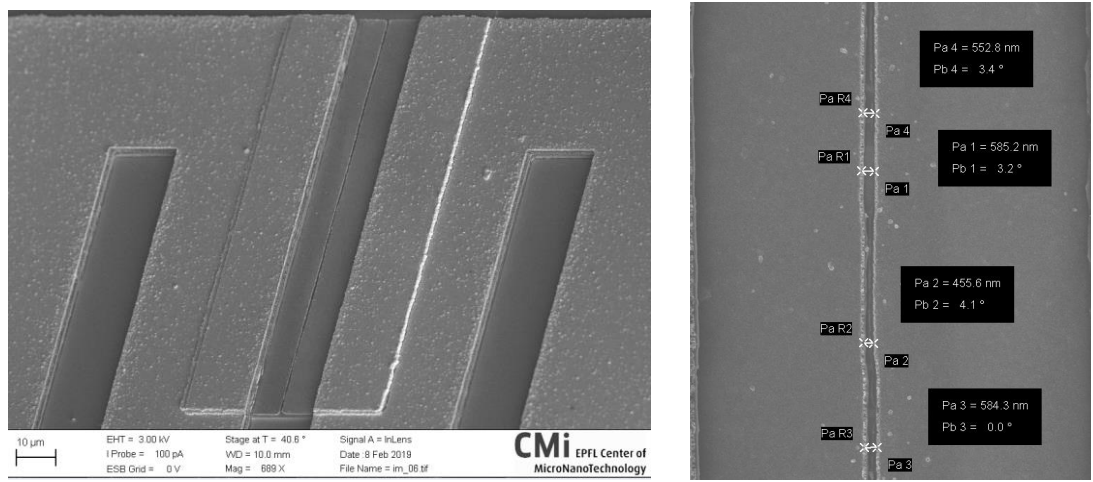

**Supplementary Fig. 22**: Peano fabricated inductor with a sub 600 nm gap in the Al connecting the VO$_2$ thin film. **a,** Inductor switch fabricated layout. **b,** Zoom of the switch ( gap within the metallization where VO$_2$ conductivity plays a key role)-in on state this gap plays a very important role-since it affects the losses ( the VO$_2$ conductivity being below 40.000S/m) on SiO$_2$/Si substrates. In off state the 20S/m conductivity of VO$_2$ adds to dielectric losses of this material.

## S4. Frequency dependence of the relative effective permittivity of VO$_2$

There relative effective permittivity (or dielectric function) $\varepsilon_{reff}$ of a material is defined as (S48) while the effective dielectric permittivity as (S49) where $\varepsilon_0$ represents the permittivity of the free space[40]. On the other hand the imaginary part of the $\varepsilon_{reff}$ defined as $\varepsilon_{reff}''(f)$ has two different components as described in (S50)-one generated by conductive losses $\sigma'(f)$ and the other related to the intrinsic tangent losses embodied in $e_r''(f)$.

$$\varepsilon_{reff}(f) = \varepsilon_{reff}'(f) - j\varepsilon_{reff}''(f) \qquad (S48)$$

$$\varepsilon_{eff}(f) = \varepsilon_0 * \varepsilon_{reff}(f) = \varepsilon_{eff}'(f) - j\varepsilon_{eff}''(f) \qquad (S49)$$

$$\varepsilon_{reff}''(f) = e_r''(f) + \frac{\sigma'(f)}{2\pi\varepsilon_0} \qquad (S50)$$

The Table I presents the variety of reported values for the relative effective permittivity real part (dielectric constant) and imaginary part as reported throughout literature[41-46]. Most studies were however done in the optical frequency range-the values and models used being not always convergent. The only study in the microwaves frequency range is done on a VO$_2$ deposition on Sapphire and are presented in[42] being based on a conformal mapping procedure which is inaccurate for the low frequency range among others (comments on it in section VI). On the other hand the model in [42] neglects the first term in (S50) attributing all the losses to a frequency dependent conductivity while ignoring the intrinsic dielectric losses. Ignoring the first term in (S50) is done in the metal modellings[40], while neglecting the dielectric damping loss. On the other hand in HFSS EM tool both terms in the right side of (S50) can be introduced unlike other tools where $\varepsilon_{reff}''$ alone exists. Thus in our extraction we used the more complex model based on (S48)-(S50)[40] while using HFSS tool which allows this separation of losses and thus a better de-embedding.

Table 1: Effective relative permittivity of $VO_2$

| Reference | $\varepsilon'_{reff}(f)$ | $\varepsilon'_{reff}(f)$ monotony | $\varepsilon''_{reff}(f)$ | Frequency | T | Observation |
|---|---|---|---|---|---|---|
| 41 | $-2 \leq \varepsilon'_{reff} \leq 10$ | r↑ | included | Optical range | $20^0C$ | |
| 42 | $400 \geq \varepsilon'_{reff} \geq 90$ | r↓ | $\varepsilon''_{reff}(f) = \dfrac{\sigma'(f)}{2\pi\varepsilon_0}$ | 0.1GHz-40 GHz | $20^0C$ | Conformal mapping Microwave range: 1-30 GHz |
| 43 | 9 | constant | - | 5.07-9.01 THz | - | |
| 34,44 | 30 | constant | - | 1 GHz-40GHz | $20^0C$ | |
| 45 | $9 \geq \varepsilon'_{reff} \geq 25$ | ↑↓ | | 76GHz-109GHz | $58^0C$ | Upper microwave spectrum |
| 46 | $7 \leq \varepsilon'_{reff} \leq 8$ | $\varepsilon_r$↑↓ | | Optical range | Insulating state | |
| 47 | 90 | constant | included | 0.1 KHz-0.1MHz | $20^0C$ | Low frequency range |

Here based on Peano filters we propose a different methodology for extracting the effective relative permittivity. Based on the Peano space filling curve we produce a variety of filters (occupying a limited area) and resonating at a variety of frequencies, additionally we consider other split ring filters too. **Supplementary Fig 23 a** illustrates the substrate layer configuration for the filters fabricated without the $VO_2$ thin film below the metallization. Fig. **Supplementary Fig 23 b** illustrates the substrate layer configuration when the $VO_2$ layer was inserted. The procedure: the defected ground plane structures (DGS) within the coplanar waveguides (CPW) resonate at a different frequency, depending on their geometrical dimensions and substrate configuration. The dielectric constant of the $VO_2$ is extracted from de-embedding the S parameters for each structures in a bandwidth next to their resonances. Each identical filter is fabricated on the substrate layer configuration from **Supplementary Fig 23 a** and then on the substrate layer configuration **Supplementary Fig 23 b**. **Supplementary Figs 24 a- Supplementary Figs 31 a-** show the different filters geometries, while **Supplementary Figs 24 b- Supplementary Figs 31 b** show the transmission parameters $S_{21}$ magnitudes in dB for the corresponding layouts in both substrates configurations.

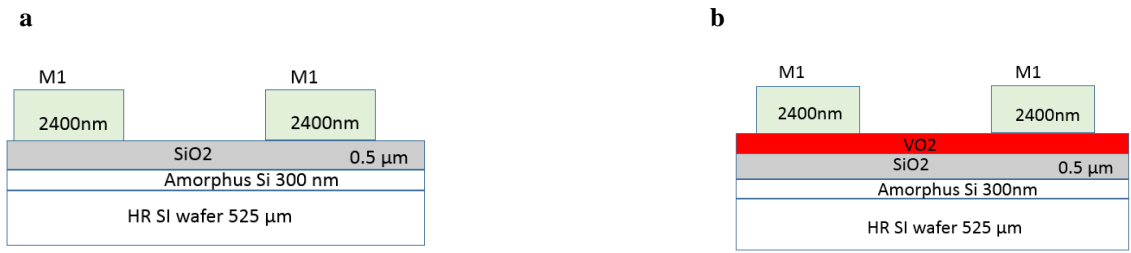

**Supplementary Fig. 23**: The fabrication of different wafers with or without a $VO_2$ layer below the metallization. **a,** the DGS CPW metallization is on $SiO_2$/Si substrate, **b,** a $VO_2$ layer is deposited below the DGS CPW metallization.

a                                                                                                 b

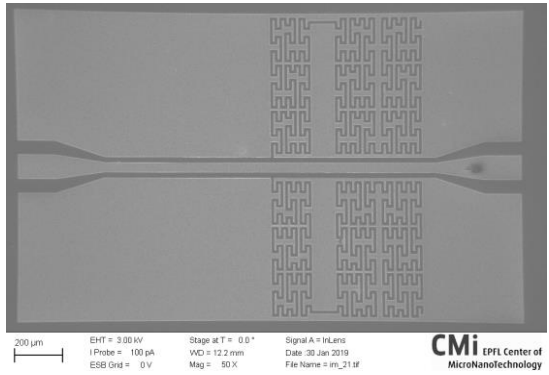
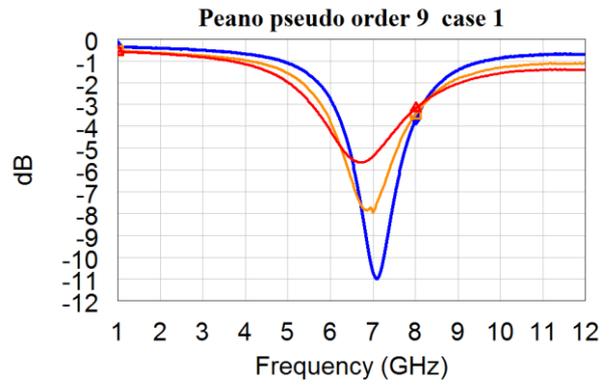

**Supplementary Fig. 24**: CPW DGS Peano filter of pseudo order 9 (9 times order 2-modified) fabricated on a SiO$_2$/Si substrate and on a VO$_2$/SiO$_2$/Si substrate. **a**, Layout of the filter. **b** Measured S$_{21}$ (dB) at 20°C: blue on SiO$_2$/Si substrate, orange with a 140 nm VO$_2$ layer below the DGS CPW metallization while red with a 170 nm VO$_2$ layer below the metallization.

**a**

**b**

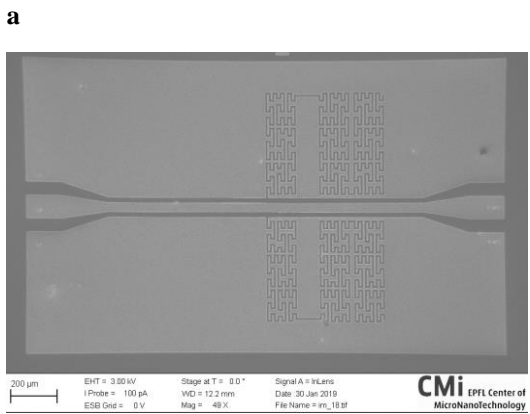
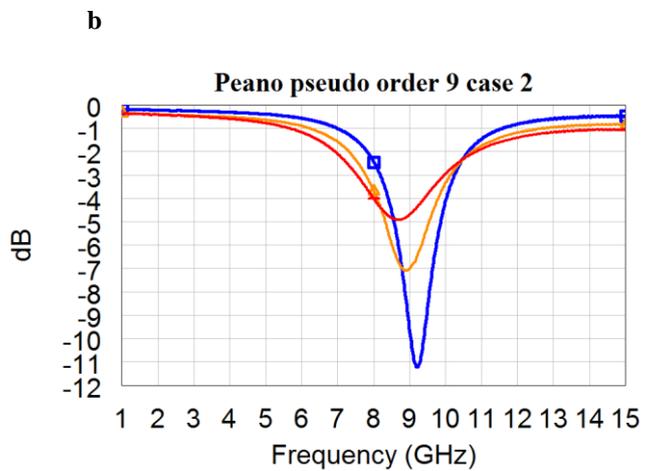

**Supplementary Fig. 25**: CPW DGS Peano filter of pseudo order 9 (9 times order 2-modified) with a different length of the Peano generating curve fabricated on a SiO$_2$/Si substrate and on a VO$_2$/SiO$_2$/Si substrate. **a**, Layout of the filter. **b,** Measured S$_{21}$ (dB) at 20°C: blue on a SiO$_2$/Si substrate, orange on a VO$_2$/SiO$_2$/Si substrate with a 140 nm VO$_2$ layer while red with a 170 nm layer. As the thickness of VO$_2$ increases the resonance frequency shift and losses are more pronounced than in respect to the same filter fabricated without the VO$_2$ layer.

**a**

**b**

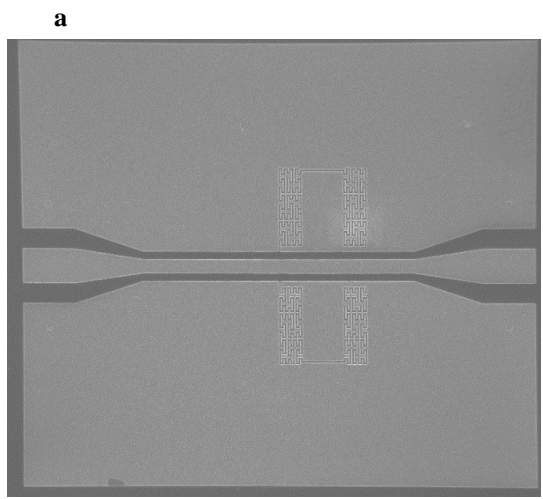
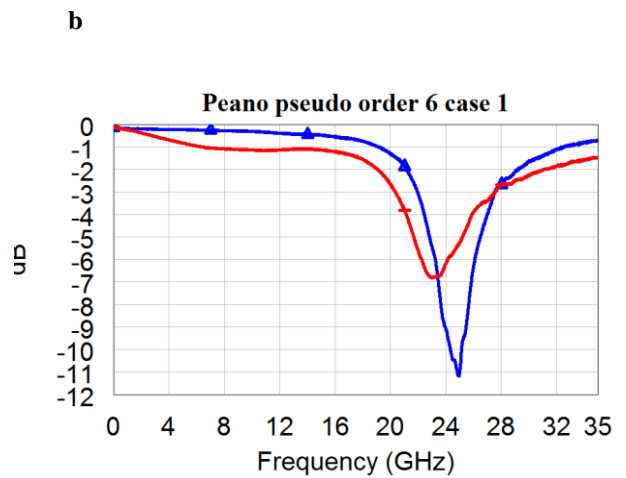

**Supplementary Fig. 26**: DGS CPW Peano filter of pseudo order 6 (6 times order 2-modified) on a SiO$_2$/Si substrate and on a VO$_2$/SiO$_2$/Si substrate. **a**, Layout of the filter. **b**, Measured S$_{21}$ (dB) at 20°C: blue on SiO$_2$ blue on a SiO$_2$/Si substrate, while red on a on a VO$_2$/SiO$_2$/Si substrate with a 170 nm VO$_2$ thick layer.

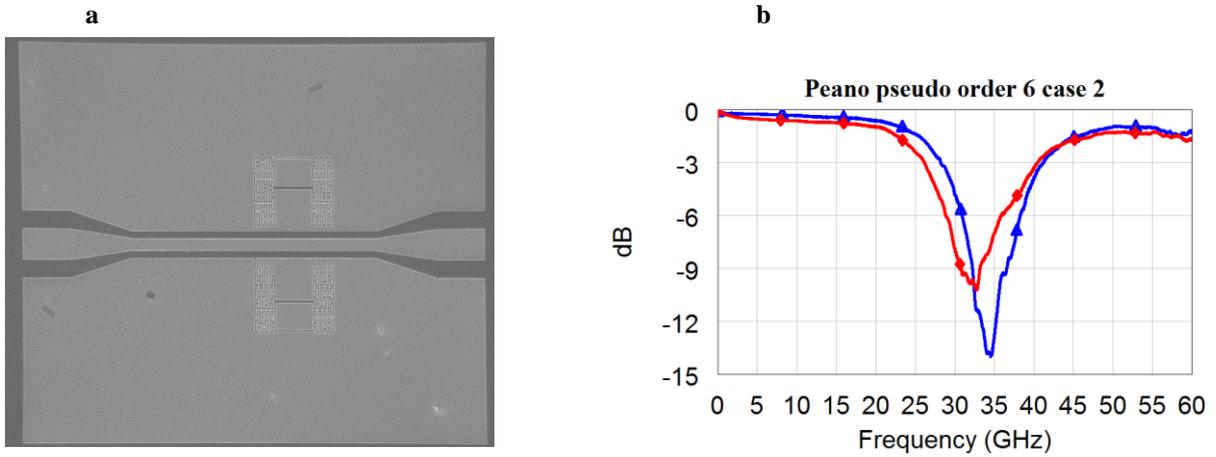

**Supplementary Fig. 27**: DGS CPW Peano filter of pseudo order 6 (6 times order 2-modified) with an additional cut. **a**, Layout of the filter. **b,** Measured S$_{21}$ (dB) at 20°C: on a SiO$_2$/Si substrate, while red on a on a VO$_2$/SiO$_2$/Si substrate with a 170 nm VO$_2$ layer.

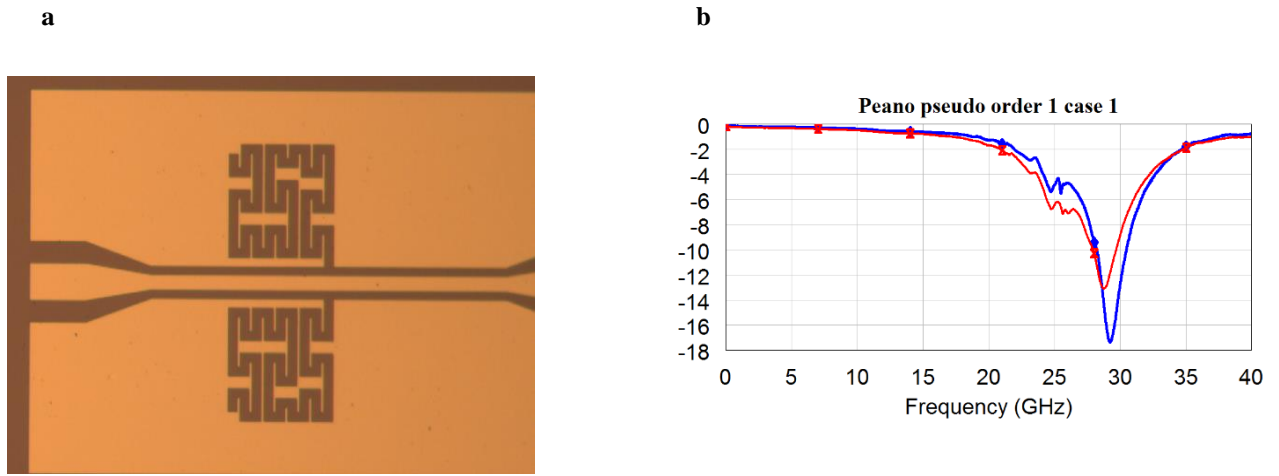

**Supplementary Fig. 28**: DGS CPW Peano filter of pseudo order 2 (order 2 slightly modified). **a**, Layout of the filter. **b,** Measured S$_{21}$ (dB) at 20°C: on a SiO$_2$/Si substrate, while red on a on a VO$_2$/SiO$_2$/Si substrate with a 170 nm thick VO$_2$ layer.

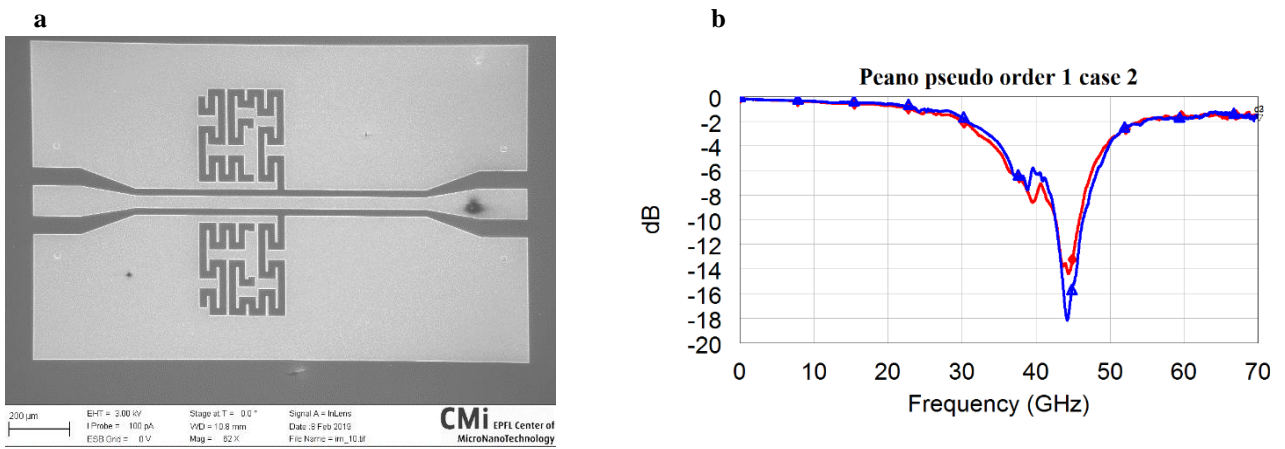

**Supplementary Fig. 29**: DGS CPW Peano filter of pseudo order 2 with a metal block on its DGS-. **a**, Layout of the filter. **b,** Measured S$_{21}$ (dB) at 20°C: on a SiO$_2$/Si substrate, while red on a on a VO$_2$/SiO$_2$/Si substrate with a 170 nm thick VO$_2$ layer .

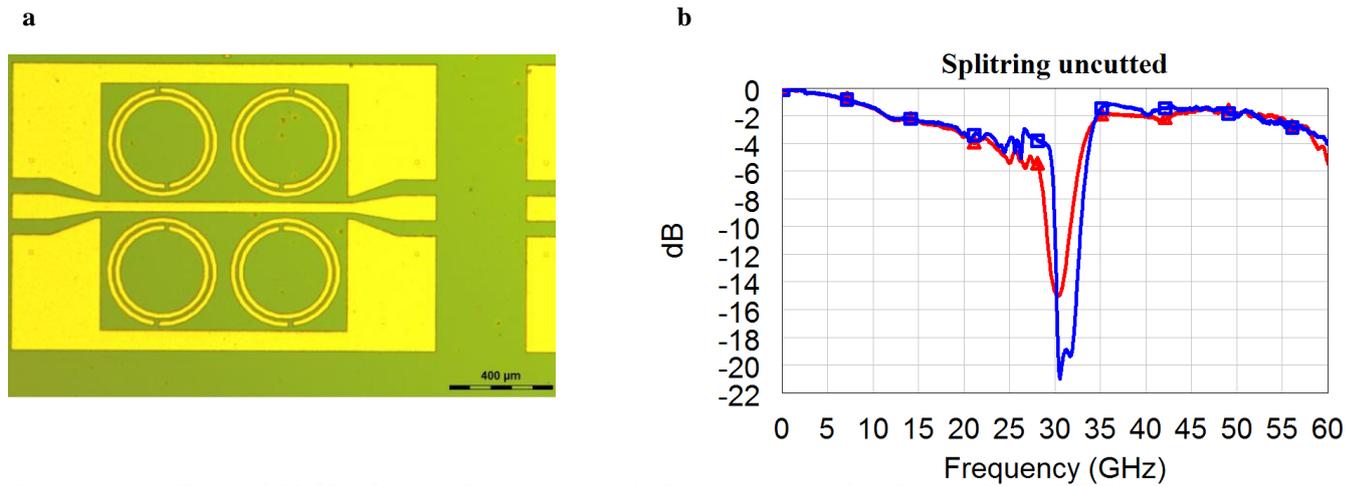

**Supplementary Fig. 30**: DGS CPW Split ring filter. **a,** Layout of the filter. **b,** Measured S$_{21}$ (dB) at 20°C: on a SiO$_2$/Si substrate, while red on a on a VO$_2$/SiO$_2$/Si substrate with a 170 nm thick VO$_2$ layer.

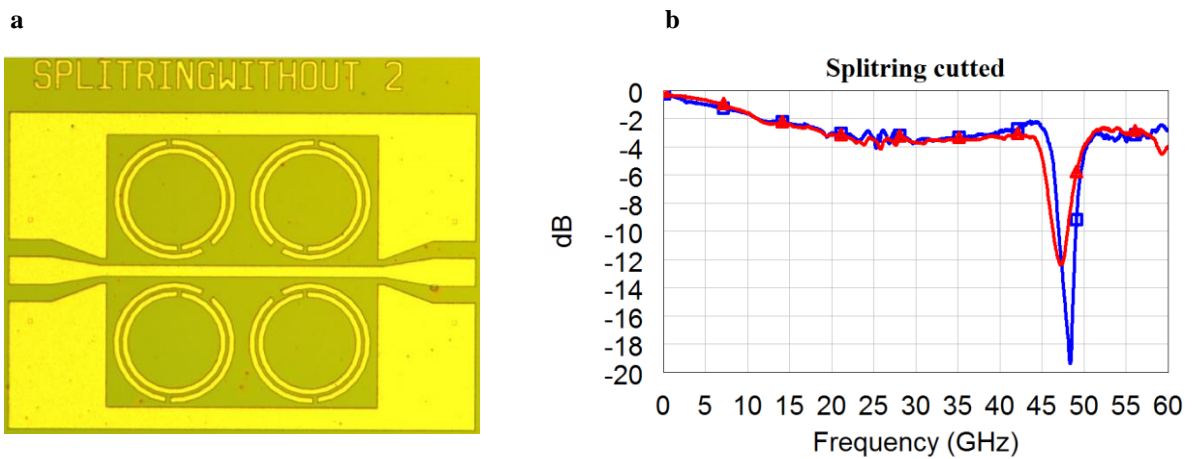

**Supplementary Fig. 31**: Split ring CPW DGS filter with an additional cut. **a,** Layout of the filter. **b,** Measured S$_{21}$ (dB) at 20°C: on a SiO$_2$/Si substrate, while red on a on a VO$_2$/SiO$_2$/Si substrate with a 170 nm thick VO$_2$ layer.

The results show clearly a lower relative frequency shift (inserting the VO$_2$ layer below the filters) as their resonant frequency increases. Nevertheless this is dependent on each geometry, each being more or less affected by the thin VO$_2$ layer. The de-embedded results using HFSS conclude to the following dielectric constant curve with the losses in **Supplementary Fig. 32**. The results are in accordance with[47] for the low frequency range while the values computed at 50 GHz are in accordance too with the ones reported in [45] for 75 GHz.

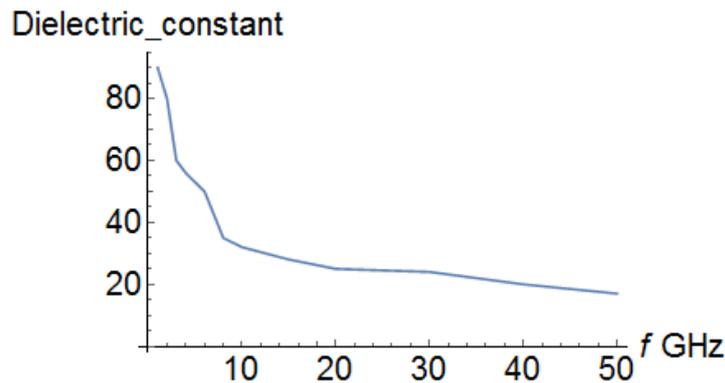

**Supplementary Fig. 32**: Extracted dielectric constant of $VO_2$ on a $SiO_2$/Si substrate at room temperature.

A supplementary video that showcases the new functionalities of the 3D Smith Chart (http://www.3dsmithchart.com/) developed specifically for the new notions presented in the article can be found at the following link:

https://www.youtube.com/watch?v=kk1aGb8d_rg